\documentclass[preprint,12pt]{elsarticle}

\usepackage{amssymb}
\newcounter{qcounter}
 \usepackage[table]{xcolor} 

\usepackage{pdfpages}
\usepackage{pdflscape}

\usepackage{lipsum}
\usepackage{blindtext}
\usepackage{multirow}
\usepackage{comment}
\usepackage{url}
\usepackage{graphicx}
\usepackage{adjustbox}
\usepackage{array}
\usepackage{booktabs}
\usepackage{tabularx}

\journal{Computers \& Security}

\begin{document}

\begin{frontmatter}



\title{Risk Assessment of Cyber Attacks on Telemetry Enabled Cardiac Implantable Electronic Devices (CIED)}


\author[label1,label3] {Mikaela  Ngambo\'e} 
\author[label1,label3] {Paul Berthier M.SC.A}
\author[label1,label3] {Nader Ammari M.SC.A}
\author[label2,label3] {Katia Dyrda MD MSc}
\author[label1,label3] {Jos\'e M. Fernandez PhD}

\address[label1] {\'Ecole Polytechnique de Montr\'eal}
\address[label2] { Montr\'eal Heart Institute, Universit\'e de Montr\'eal}
\address[label3] {Qu\'ebec, Canada}

\begin{abstract}
%\paragraph*{Background}

\noindent Cardiac Implantable Electronic Devices (CIED) are fast becoming a fundamental tool of advanced medical technology and a key instrument in saving lives. Despite their importance, previous studies have shown that CIED are not completely secure against cyber attacks and especially those who are exploiting their Radio Frequency (RF) communication interfaces. Furthermore, the telemetry capabilities and IP connectivity of the external devices interacting with the CIED are creating other entry points that may be used by attackers. Although the majority of these vulnerabilities are more like proof of concepts or in-the-lab experiments and that there are no indicators of active exploitation or in the wild abuse, it remains crucial to perform a risk analysis to measure how viable these attacks are, their impact and consequently the risk exposure. In this paper, we carry out a realistic risk analysis of such attacks.  This analysis is composed of three parts. First, an actor-based analysis to determine the impact of the attacks. Second, a scenario-based analysis to determine the probability of occurrence of each threat. Finally, a combined analysis to determine which attack outcomes (i.e.~attack goals) are riskiest and to identify the vulnerabilities that constitute the highest overall risk exposure. The conducted study showed that the vulnerabilities associated with the RF  interface of CIED represent an acceptable risk. In contrast, the network and internet connectivity of external devices represent an important potential risk. The previously described findings suggest that the highest risk is associated with external systems and not the CIED itself. A noteworthy observation that emerged from the risk analysis is the fact that the damages of these cyber attacks could spread further to affect parties other than patients such as device manufacturers through intellectual property theft or medical practitioners through affecting their reputation. 
This research work has contributed to extend the knowledge in terms of quantifying the risk associated not only to CIED devices but also to their ecosystem.  The results of this study could be considered as a base for CIED risk management procedures as they help to measure the impact of different attacks while taking into consideration the attackers goals, identifying attack scenarios as well as their likelihood of occurrence and determining which threat has to be addressed in priority.

\end{abstract}


\end{frontmatter}

%\linenumbers

\section{Introduction}

\noindent Cardiac implantable electronic devices (CIED) have evolved from single-chamber pacing devices to  resynchronization and defibrillation within the same device \cite{LibroKatia} . Modern CIED now include numerous functionalities being integrated into a single device, which has contributed to an increase in the number of implanted devices\cite{savci2005mics, sanders1996implantable}.  Besides, the use of telemetry-enabled CIED is increasing at the detriment of older models with no wireless-communication capabilities \cite{Eluna,CARELINK}, due to the significant advantages it brings to patient care \cite{slotwiner2015hrs,ricci2013effectiveness}.  For the remainder of this article, the acronym CIED will refer only to telemetry-enabled CIED.\\

\noindent CIED interact with external systems located in the hospital (the \emph{external   programmer}), the patient's home (the \emph{home monitor}) and in the cloud \cite{CARELINK,EcosystemER,homeMonitor}. They communicate with the external programmer and the home-monitoring device via Radio Frequency (RF) signals transmitted in the Medical Implants Communication Services band (MICS 402-405 Mhz) \cite{savci2005mics,islam2016review,federal1999medical,rules2003regulations, mics}, whereas they interact with cloud-based systems by means of the home-monitoring device and Internet Protocol (IP) connectivity \cite{slotwiner2015hrs,ricci2013effectiveness,homeMonitor}.\\
\newpage

\noindent External programmers are used by the physicians \emph{ab initio} when configuring the devices prior to implantation and during patient follow-up sessions to retrieve data and for reconfiguration. They have three modes of operation: 1) the \emph{interrogation mode} to check a patient's cardiac programmed parameters and stored data, 2) the \emph{test mode} to test that the implant is operating properly, and 3) the \emph{programming mode} which allows the physician to adjust the patient's therapy by reconfiguring the functionality of the CIED \cite{Eluna,CARELINK}.  Reconfiguration after manufacturing is feasible since current CIED circuitry is microprocessor-controlled and its software can be updated \cite{sanders1996implantable,bernstein1983microcomputer,stotts1989vlsi,baker1985pacemaker,sholder1991programmable,brockway1986programmable,duggan1990adaptable,segerstad1984pacemaker}. \\

\noindent Home monitoring devices are intended to supervise the patient's cardiac status. They periodically collect activity data from the CIED and send it to a cloud-based database. The latter may be operated by either the CIED manufacturing company or a web services provider used by the physician to access the patient's data \cite{slotwiner2015hrs,ricci2013effectiveness,homeMonitor}.\\

\noindent  As evidenced by previous work, CIED are vulnerable to cyber attacks that use their RF interfaces to communicate with the devices \cite{halperin2008pacemakers,marin2016security}. This is also true for non-telemetry enabled CIED, but telemetry introduces additional vectors of cyber attacks that can include manipulation of the home monitor, interception of transmissions from the home monitor to the cloud and the physician's station, and manipulation of the cloud-based database itself \cite{EcosystemER, burns2016brief}.  Proof of the increased concern of cyber attacks on CIED was given by the recall of almost half a million CIED by the Food and Drug Administration (FDA) in August 2017. According to the FDA, the aforementioned devices were vulnerable to unauthorized access, allowing a malicious person to reprogram them using commercially available equipment \cite{FDA}. However, no such attacks have been reported. While we know it would be technically possible to conduct such an attack in the controlled environment of a research laboratory \cite{halperin2008pacemakers, marin2016security,EcosystemER}, it remains to be determined how viable such an attack would be on an actual target in the real world. This is precisely our research question: What are the real-life risks of cyber attack onto telemetry-enabled CIED and the systems they depend on?\\

\newpage

\noindent In this work, we carry out a realistic risk analysis of such attacks, with regards to actual impact these problems pose in terms of: health, economy, quality of life and privacy of the affected parties. In order to carry this risk analysis, we inventory the vulnerabilities that have been made public up to now, we define attack scenarios based on them, describe and evaluate the impact of the various attack goals that various actors would want to achieve through such attack scenarios, and finally estimate the likelihood of occurrence of each of these attacks to determine overall risk. \\

\noindent Our motivation to conduct this research is based on the need to understand the real scope of the problem.  After the FDA statement was released, patients began to massively call their cardiologists to get an explanation about these potential failures and to what extent they were in danger. It is at times difficult for physicians to answer them, since cybersecurity is not their field of expertise and because there is little information about the clinical impact of exploitation of the vulnerabilities found. This is why we believe that such a ``reality check'' is necessary, as the real scope of the problem it is not clear at all. By determining the scope of the problem we contribute to 1) extend the knowledge of the threats affecting CIED,  2) provide guidance on which threats should be addressed in priority and consequently 3) provide to the organizations potentially interested in this kind of risk assessment a basis from where to start, e.g.~health regulation agencies, device manufacturer, health practitioners, etc.\\

\section{Background on CIED}
\subsection{CIED computer-based architecture}

\begin{figure}[ht!]
\begin{center}
  \includegraphics[width=0.8\linewidth]{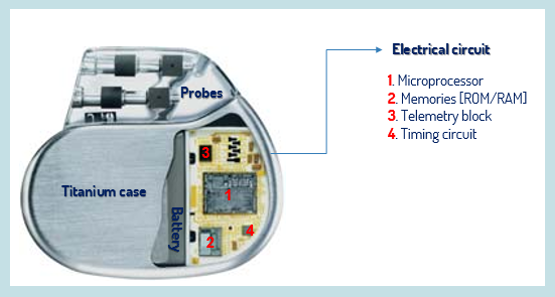}
  \caption{CIED circuitry}
  \label{fig:circuit}
\end{center}
\end{figure}

\noindent Few documentary sources on CIED architecture and design are available, due to the proprietary nature of that information. Nonetheless, common principles and some technical details are generally documented in some of the corresponding patents \cite{sanders1996implantable,baker1985pacemaker,sholder1991programmable,brockway1986programmable,duggan1990adaptable,segerstad1984pacemaker}. As depicted in Figure~\ref{fig:circuit}, today's software-based CIED circuitry is mainly composed of four electronic components with the first one  being the \textit{microprocessor} which is the ``brain'' of the CIED. It coordinates, controls and directs the interactions between the elements of the circuit. It also interprets and executes the algorithms programmed in memory. The second component is the \textit{memory}, CIED holds two kinds of memories: Read-Only Memory (ROM) and Random-Access Memory (RAM). The ROM contains the embedded software (also known as \emph{firmware}) providing low-level control of the device's hardware, as well as the code implementing the various functionalities of the device. While the RAM is taking care of storing a variety of parameters, such as device serial number, patient ID, clinical information, patient's cardiac activity (arrhythmia logs, frequency histograms) and certain programs implementing particular therapies. The third component dubbed the \textit{telemetry circuitry} is used to establish a communication link between the CIED%\footnote{In this dissertation, we refer to the CIED as a pulse generator or pacemaker. } 
and external devices, such as the external programmer or a home monitor. More specifically, the telemetry circuitry allows performing remote monitoring, therapy adjustment and reprogramming the CIED prior to implantation or during patient follow-up sessions. The fourth electrical component is the \textit{timing circuit} which is a key element on the CIED circuitry as it takes care of synchronizing the stimulation pulses to the cardiac chambers as well as the memory access.\\

\begin{figure}[]
\begin{center}
  \includegraphics[width=0.9\linewidth]{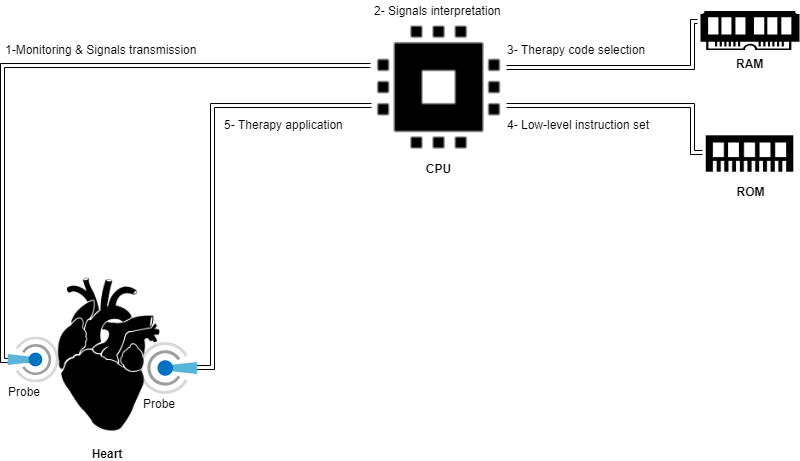}
  \caption{Therapy selection loop}
  \label{fig:TherapySelection}
\end{center}
\end{figure}
\noindent The previously described components are interacting together to maintain the CIED functionalities, each one of them has a specific task that has to be executed in a coordinated way. A good example of these multi-component interactions would be the automatic therapy selection loop (Figure \ref{fig:TherapySelection}). This process is performed at the microprocessor level using signals coming from the detection probes that are continuously monitoring and re-transmitting the cardiac activity to the microprocessor. The re-transmitted signals are then interpreted in order to select the adequate therapy code from the RAM. Finally, the selected code is translated to a low-level instruction set (located in the ROM) before being executed by the microprocessor. It is also important to mention that the treating physician is responsible of determining which therapies can be applied under what conditions.\\
 
\begin{figure}[]
\begin{center}
  \includegraphics[width=0.5\linewidth]{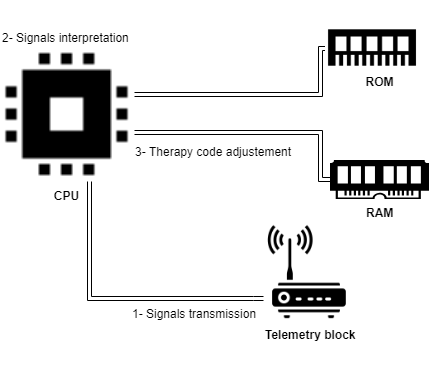}
  \caption{Therapy adjustment}
  \label{fig:TherapyAdjustement}
\end{center}
\end{figure}
\noindent The clinician uses the programmer's user interface to program or adjust the parameters of the patient therapy (i.e., number of beats per second).  When fixing a therapy parameter on the programmer user interface, the external device subsequently sends to the CIED an instruction containing the change to perform on the therapy. The telemetry circuitry receives this instruction and then sent it to the microprocessor. Following the interpretation of the instruction, the microprocessor will access the RAM to perform the required change. (Figure \ref{fig:TherapyAdjustement}).\\

\subsection{CIED ecosystem}
\begin{figure}[] 
\includegraphics[width=0.8\linewidth]{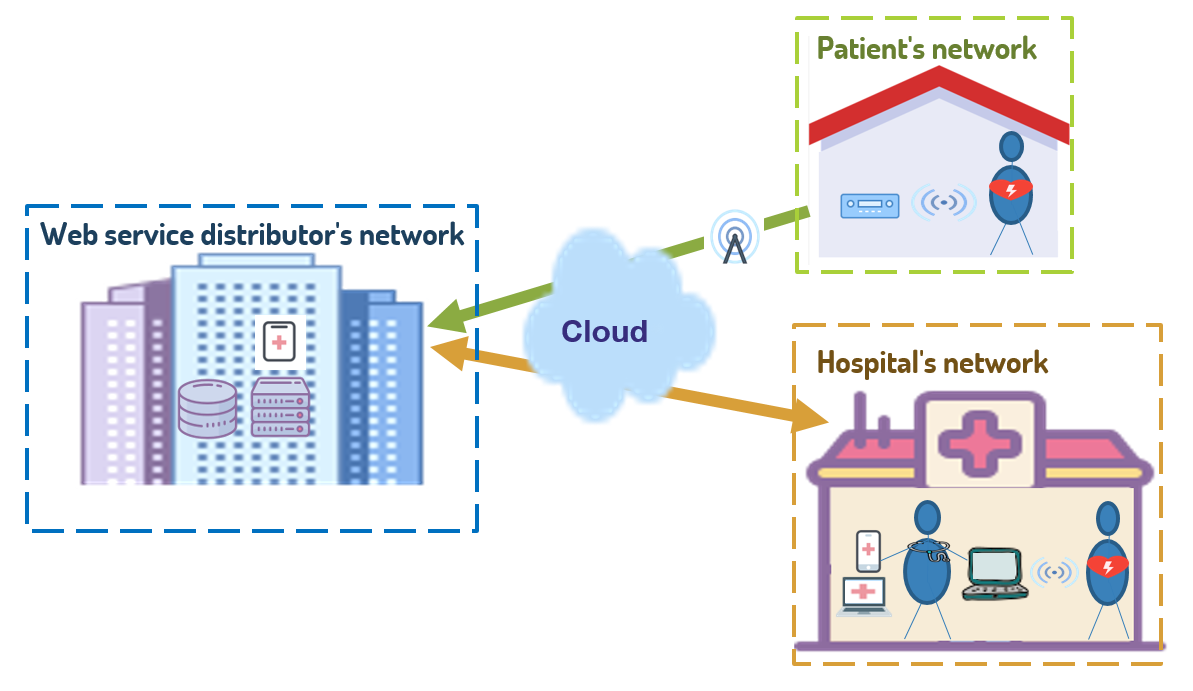}
\caption{CIED's ecosystem.}
\label{fig:ecosystem}
\end{figure}

\noindent The CIED ecosystem (Figure \ref{fig:ecosystem}) encompasses the set of devices,  cloud-based systems and cloud-based services employed for the diagnosis, the therapy's adjustment and the monitoring of patients with an implanted CIED. Apart from the CIED itself, there are two other medical devices forming part of this ecosystem. These are, the external programmer usually located at the hospital and the monitor located at the patient's home. Health professionals rely on the external programmer to obtain the programmed parameters of the patient, to adjust the desired therapies or to check the correct operation of the CIED \cite{Eluna,CARELINK}. The home monitor is used to periodically collect the data stored in the CIED and send them to a cloud-based database (DB). Thus, medical staff can access a patient's health information through a web-based application, operated either by the CIED manufacturer or a separate cloud service provider \cite{slotwiner2015hrs,ricci2013effectiveness,homeMonitor}. It is fair to say that the monitor is a key element of this system of systems as it is through him that some CIED data are available on the cloud-based elements of the ecosystem in study. By cloud-based elements, we refer to the cloud-based database containing the information coming from the monitor, the cloud-based application displaying this information and, the cloud-based server containing the application. Moreover, the set of devices employed to use the application mentioned above (e.g. smart-phones, tablets, and laptops) form also part of the ecosystem of CIED. \\

\noindent The CIED interact with all the elements of their ecosystem. Depending on the element, the interaction could be either direct or indirect. Direct interaction takes place with both the external programmer at the hospital and the home monitor device, while indirect interaction occurs with the cloud-based systems and services. We distinguish the type of interaction since as we will see later (Section \ref{AttackScenario}), it determines the kind of attacks that can be carried out. \\

\noindent \textbf{Direct interaction} consists of wireless communication between the CIED and a programmer or a home monitor device. Indeed, current pacing devices include two types of wireless technology.  The first, referred to as \emph{Inductive-coil telemetry}, uses an inductive RF field (0-300 kHz) to communicate over short ranges (0-10 cm), requiring proximity between the CIED and the antenna coil of the external programs. The second mode referred to as \emph{RF-link telemetry}, uses a radiating RF field (i.e.~traditional RF waves) at a higher frequency (402-405 MHz) to communicate over longer ranges (0-200 m) \cite{EcoRFtypos}.  Both technologies can be used for interrogation and programming operations during follow-up visits at the hospital.  The main difference between them is that with inductive coil telemetry it is necessary to apply a programming head\footnote{A programming head is an  sub-component of the programmer serving as a security switch that can activate or deactivate data transmission to/from the CIED. } right above the CIED to establish communications, while with RF link this can be achieved without such proximity; it is therefore referred to as ``wand-less" \cite{Eluna,CARELINK}. Nowadays, the trend is to use RF-link telemetry to the detriment of the inductive-coil telemetry, since it has a better conductivity in the human body, a higher data rate, a greater communication range, and thus constitutes a more adequate option for home monitoring \cite{Eluna,CARELINK}. This technology operates on the Medical Implant Communications Service (MICS) core band (402-405 MHz) initially allocated by the Federal Communication Commission (FCC) in 1999 \cite{rules2003regulations}, and that has subsequently been adopted in different regions of the world \cite{islam2016review,savci2005mics}. \\

\noindent The MICS spectrum was established in order to support data transmission between any implanted and external device for diagnostic and therapeutic purposes, not only for CIED \cite{islam2016review,mics}.  However, since this band is shared with communication devices used by meteorological services, its use has been normalized to avoid any interference \cite{savci2005mics,federal1999medical,rules2003regulations}.  These rules of use are based on ITU-R Recommendation\footnote{The ITU-R Recommendations are  international technical standards elaborated by the Radiocommunication sector of the International Telecommunication Union.} SA.1346 \cite{ITU-R}, whose broad outline is the following:
\begin{itemize}
\item The MICS spectrum is divided into 10 transmission channels with a   bandwidth of 300 kHz each.
\item The implanted devices can only transmit after receiving a command from an external device authorizing them to do so (there are exceptions such as emergencies).
\item External devices must integrate interference mitigation techniques, such as Listen Before Talk (LBT) and Adaptive Frequency Agility (AFA) to minimize the effect of ambient noise. External device radio transmitters use LBT to sense their radio environment before starting a transmission, in order to identify a free channel with least interference, i.e.~the Least Interference   Channel (LIC). Channel verification is done five seconds before initiating a   transmission, with each channel being monitored for 10 ms to determine if it is occupied. Then, based on the sensed information the transmitters use the   AFA technique to select an operating frequency that is not in use yet   \cite{LBT,AFA}.
\end{itemize} 

\noindent Although the use of the transmission means adheres to an internationally-adopted standard and is carefully regulated, the same cannot be said for signal-conditioning techniques or CIED access control mechanisms. While the Health Insurance Portability and Accountability Act (HIPAA)\footnote{The HIPAA was created in 1996 to enhance the performance of the U.S healthcare system.  The progressive adoption of IT  in the health services has led to the inclusion of security and privacy rules on the HIPPA. Those rules protect the sensitive data of the patients that are electronically stored or transmitted .}determines privacy and security rules for the protection of medical data transmitted between systems, these remain very general and are more oriented towards traditional Information Technology (IT) systems such as computers and servers. While the FDA has defined good practice guidelines in this area, they do not \emph{establish legally enforceable responsibilities} \cite{FDApostmarketSurveillance, MIT2015improving} so their application remains optional. This legislative vacuum results in an overabundance of proprietary communication and authentication protocols, where it is up to each manufacturer to set their own criteria and choose the security methods to apply to their devices.\\

\noindent \textbf{Indirect interaction} occurs between the CIED and the cloud-based systems or services when those are employed to display the information contained in the CIED. Health practitioners have the ability to access the collected data using their own portable devices via connection to the cloud-based web application through a regular Internet connection. This approach aims to improve the quality of patient care and the working conditions of medical staff while reducing health costs. Thus, even if strictly speaking there is no direct communication between the CIED and these systems, there is still a data link between them. Nonetheless, current CIED features do not include the capability to download information from the DB server, and in principle cannot be remotely reconfigured by the health practitioners from the monitors. \\

\noindent In summary, CIED operate in a complex and heterogeneous ecosystem composed of various devices connected to different networks and using multiple communication protocols to interact with each other. The necessity to reduce health costs is propelling a worldwide transformation of health service management that relies precisely on the interaction between medical devices and health practitioners. Nevertheless, this interaction, which is undoubtedly beneficial for all stakeholders, may have turned into a significant attack surface for cyber attacks against the CIED ecosystem \cite{bsi}.\\
% AQUI
\section{Background on Implantable Medical Devices (IMD)\hbox{cybersecurity}} \label{RL}
\noindent The security threats that affect  CIED apply to all IMD. The latter are vulnerable to malicious exploitation of  (i) their RF communication interfaces, and (ii) the telemetry functionalities and IP connectivity of the extracorporeal equipment on which they depend. In this section, we develop a critical review of the literature on cyber threats affecting  IMD and risk assessments regarding IMD risk exposure to cyber attacks. We proceed in this way for two reasons. First, there is no abundant and exclusive literature on the cybersecurity threats of CIED. Second, we claim that the risk assessment methodology proposed here applies to all IMD.
\subsection{IMD Cyber Threats } 
\noindent In the last decade, several research groups have exposed vulnerabilities in implantable medical devices (IMD), such as insulin pumps and, of course, CIED. Below, we chronologically describe these findings.\\

\noindent In 2008, Halperin \emph{et al.}  \cite{halperin2008pacemakers} unveiled CIED vulnerabilities to radio frequency-based attacks. By making use of a software-defined radio (SDR), the researchers succeeded in reverse engineering the device's communication protocol and conducting attacks such as sensitive data interception and dangerous command emission (electric shock dispensation).\\

\noindent In 2011, Hei \emph{et al.} \cite{hei2010defending} found vulnerabilities that allow resource depletion attacks on insulin pumps. They demonstrated that by sending periodic wireless commands to the IMD, it was possible to keep an active communication session permanently opened, thus significantly reducing the device's service lifetime. During the same year, Li \emph{et al.} \cite{li2011hijacking} disclosed insulin pump vulnerabilities that allow unauthorized parties to communicate with the device. In fact, their work revealed that some insulin pump models possess a class of vulnerability that can allow an unauthorized party to ``\emph{emulate   the full functions of a remote control: wake up the insulin pump, stop/resume   the insulin injection, or immediately inject a bolus dose of insulin into the   human body}''\cite{burns2016brief}. That same year, Jerome Radcliffe, a patient with diabetes, partially reverse-engineered the communication protocols of his insulin pump.  He announced his findings at the Black Hat cybersecurity conference.\\

\noindent In 2012, the hacker Barnaby Jack demonstrated that certain CIED models can disclose their authentication credentials following the reception of a specific command \cite{ruxconbreakpoint2012}. He was also able to verify that the access codes of some devices are simply their serial and model number. Paradoxically, this information is disclosed by some CIED models when they receive a specific command from an external programmer or a home-monitoring device. This discovery highlighted that an unauthorized party could gain control of certain CIEDs by simply sending a command \cite{burns2016brief,BarnabyJackBlog}.\\

\noindent In 2016, Marin \emph{et al.} \cite{marin2016security} used a black-box reverse-engineering\footnote{Method by which an attacker discovers the structure and function of a software by interacting indirectly with it, for example through input and output vectors, libraries or APIs.} approach to analyze the proprietary communication protocols employed by CIED to communicate with external programmers over a long-range RF channel. Their work evidenced that reverse-engineering CIED proprietary communication protocols is feasible and that they present several implementation weaknesses. As proof of that, they were able to implement a set of exploits like the interception of sensitive information, Denial-of-Service (DoS), spoofing and replay attacks. The findings of this research were reproduced for at least 10 different models of CIED.\\

\subsection{IMD cybersecurity risk analysis }
\noindent Even though the first study evidencing vulnerabilities in CIED was published in 2008 \cite{halperin2008pacemakers}, it is only in 2015 i.e seven years later that the first IMD risk assessment study appears in the literature.\\

\noindent In 2015  Jagannathan and Sorini conducted a full IMD-specific cybersecurity risk analysis \cite{cybersecurity}.  This study presents a methodology to evaluate medical devices exposure to cybersecurity risks. The method presented is a traditional Preliminary Hazards Analysis (PHA) study which was tailored to assess the cybersecurity properties of medical equipment. As any PHA study, their methodology consists of three main steps namely: 1) hazard identification, 2) risk determination, and 3) risk ranking and follow-up actions. This work analyses the cybersecurity risk of fictitious medical devices. Thus, its findings do not reflect the actual state of the problem. In our work, we analyze real medical devices that are currently in the market. Therefore, the results herein find not only illustrate the actual scope of the problem but can serve as a basis for the risk management procedures related to the CIED ecosystem.\\

\noindent In April 2017, a study by Stine \emph{et al.}  \cite{CyberiskScore} presented a cybersecurity risk assessment method for network-connected medical devices. This study introduced a scoring system relying on a cybersecurity questionnaire based on the STRIDE\footnote{STRIDE is the acronyms of a set of computer attacks namely Spoofing, Tampering, Repudiation, Information Disclosure, Denial of Service, Elevation of Privilege} model developed by Microsoft for classifying threats \cite{STRIDE}. The scoring system is intended to help healthcare organizations in identifying those medical devices that have the potential to endanger patient health or disrupt the quality of medical follow-up. This study estimates the probability of occurrence of an attack according to the security features implemented in the target system. Since the estimate of P  is only based on the technical difficulty of the attacks, it does not adequately reflect reality.  Just because an attack is technically simple to carry out does not mean that an attacker will be interested in achieving it. The chance and willingness to attack are essential factors when it comes to estimating P.  Accordingly,  in our study, we estimated P in function not only of the characteristics of the target system but also,  in function of specific characteristics of the attacker.\\

\noindent In May 2017, Rios and Butts \cite{EcosystemER} conducted an exhaustive analysis of the CIED ecosystem and the interdependence between its elements. The hardware and software components of different models of CIED, external programmers and home-monitoring devices from different manufacturers were examined. As a result, over 8,000 known vulnerabilities were discovered in third-party libraries of four external programmer models belonging to four different manufacturers.  Besides, vulnerabilities were found in all CIED evaluated. The publication of this work preceded the massive recall of CIED ordered by the FDA in August 2017, based on the vulnerabilities reported by the Industrial Control Systems Computer Emergency Response Team (ICS-CERT) of the US Department of Homeland Security (DHS). This work identifies the threats and their nature however, the real scope of the risk that those threats entail is not described. We consider that although there are vulnerabilities in a system, it is their probability of exploitation and the impact that this exploitation has on individuals that determines whether the vulnerability represents a significant risk or not. In our study, we estimate the risk of a vulnerability based on the probability that it will be exploited and, the impact that the exploitation will have on victims.\\ %\footnote{https://ics-cert.us-cert.gov/}.

\noindent In 2018, Abrar \emph{et al.} \cite{Cloudcomp} conducted a risk analysis on cloud computing within the context of health applications, in order to evaluate their suitability for the Health Infrastructure System (HIS). The research team identified HIS vulnerabilities and then analyzed the impact that a security breach would have on its integrity if the vulnerable elements were deployed in a cloud computing environment. This paper analyzes a mortgage situation, while we analyze a real situation, i.e. the risk that current CIED cloud-based services represent for patient safety.\\

\noindent Our cybersecurity risk assessment is divided into three analysis: 1) actor-based analysis, 2) scenario-based analysis and 3) combined analysis. On the first one, we identify the potential actors and determine the impact of the attacks according to four separate aspects: health (H), monetary (M), privacy (P) and quality of life (QL). Thus, the outcomes of this work may support the objectives of different kind of organizations potentially interested in CIED risk assessment, e.g.~health regulation agencies, device manufacturer, health practitioners, etc. On the second analysis, We determine the attack scenarios and then estimate their probability of occurrence according to the actors capacity, opportunity and motivation to achieve the attacks. Finally, in the last analysis, we calculate and characterize the risk associated with each threat (actor, scenario).\\

%Finally, unlike all previous risk analysis, we measure the impact of an exploitation based on the clinical risks it involves for the patient using the Hayes classification \cite {hayes} that establishes different levels of impact on patient health, that was introduced to classify the impact of different levels of clinically significant electromagnetic interference with pacemakers.

\section{CIED ecosystem's cybersecurity risk assessment methodology}
\subsection{Aim of the risk assessment methodology}
\noindent The number of IMD that rely on ICT to ensure patients therapy, diagnosis or follow-up is increasing. However, unlike traditional ICT systems (computers, servers, networks...),  there is not yet a formal and effective method to assess the cybersecurity risk incurred by IMD. As mentioned in section \ref{RL}, even if the first vulnerabilities were found in 2008, it is only recently in 2015 that the first IMD risk assessment appears in the literature. Therefore, this area of study is just starting to develop, hence the need for a methodology.\\ 

\noindent In this work, we propose a method for asses the cybersecurity risk incurred by CIED, a subcategory of IMD. Our method can be used to guide organizations interested in conducting risk assessments of CIEDs and can be extended to other IMD.\\ 

\subsection{Definitions}\label{sectionDefinition}
\noindent We define here the cybersecurity and risk assessment terms used in this article.% in order to facilitate the reading of our work.
\par \noindent \textbf{Actor}: A person or organization that violates the integrity, privacy or confidentiality of a computer system`s data to obtain a benefit.  
\par \noindent \textbf{Impact}: Quantification of an attack's effect or consequence on the target or victim.
\par \noindent \textbf{Victim}: A person or organization that is the subject of a computer attack. 
\par \noindent \textbf{Attack goal}: Final effect desired by the actor, resulting in a negative impact on the target system or victim.
\par \noindent \textbf{Scenario}: Set of actions carried out by the actor to achieve his attack goal.
\par \noindent \textbf{Threat}: A combination of a person with deliberate intent (actor) comitting acts in particular fashion (scenario), resulting in a negative consequence (impact).
\par \noindent \textbf{Vulnerability}: A design, manufacturing or programming flaw in a system that may offer the opportunity to conduct an attack on it.
\par \noindent \textbf{Attack vector}: Subset of vulnerabilities for which there is a demonstrated attack method by which the vulnerability is employed (exploited) by the actor to reach its final goal or an intermediate goal towards it (e.g.~gaining access).
\par \noindent \textbf{Exploit}: Subset of attack vectors related to software vulnerabilities.
\par \noindent \textbf{Probability}: Likelihood that a particular threat (a given actor successfully reaching an attack goal through a given scenario) be materialized during a given period of time.
\par \noindent \textbf{Risk}: Quantification of a threat (Risk = Impact * Probability).

%\subsection{Aim of risk assessment in CIED}
%In IT, the aim of risk assessment is to evaluate risk in order to: 1) identify the riskiest threat and address them in priority, and 2) choose the most effective counter-measure by determining those with less residual risk. Nevertheless, the aim of risk analysis in our work is more extensive. Here, we evaluate risk to identify the riskiest threats (actor, scenario). Then, we identify which attack vectors represent the most cumulative risk.  Finally, we determine which are the vulnerabilities to address in priority.  We proceed in this way because, although it is true that for the traditional IT system there exist a standard ``toolbox'' of countermeasures to mitigate cyber attacks, this is not the case for CIED.  Indeed, no ready-made solution exists to choose from. Solutions must be implemented by manufacturers (security by design), possibly driven by government regulation. Thus, the ultimate goal of such a risk assessment is to provide guidance on which attack vectors and related vulnerabilities should be addressed in priority.

\begin{table*}[]
    \centering
    \resizebox{\linewidth}{!}{
    \begin{tabular}{l c l}\toprule
        \textbf{Type}&\textbf{Level}& \textbf{Description of the Impact}  \\ \midrule  
         
         Health& 1& Minor harm to the patient\\
         &2&Significant harm to the patient not involving serious life threatening injuries \\
         &3& Severe harm to the patient that involve serious life threatening injuries\\ 
         &4&Catastrophic harm to the patient that involve loss of life \\ \hline

         Monetary& 1&Minor monetary loss\\
         &2&Significant monetary loss\\
         &3 & Severe monetary loss \\
         &4& Catastrophic monetary loss \\ \hline

         Quality of life& 1&Minor impact on the patient's quality of life\\ 
         &2&Significant impact on the patient's quality of life\\
         & 3& Severe impact on the patient's quality of life\\ 
         &4& Catastrophic impact on the patient's quality of life\\ \hline

         Privacy& 1&Minor impact on the patient privacy if information is                            disclosed\\ 
         &2&Significant privacy impact if information is disclosed \\ 
         &3&Severe impact on the patient privacy if the information is disclosed \\ 
         &4& Catastrophic impact on the patient privacy if the information is disclosed\\  \hline
    \end{tabular}
}
   \caption{Impact levels}
    \label{tab:ImpactLevels}
\end{table*}

\subsection{Risk assessment methodology} \label{methodology}
Our risk assessment methodology is divided into three steps:
\begin{description}
\item[Step 1. Actor-based risk analysis] In this phase, we aim to determine and quantify the impact of attacks on the CIED ecosystem.  To do this, we first identify potential actors that would be interested in attacking the CIED ecosystem.  Then, we determine their likely attack goals and from there we quantify the impact on the victim of the successful accomplishment of such attack goals.  We do this separately according to four different categories of impact: Health, Monetary, Quality of Life and Privacy. We measure the impact on health by applying the Hayes classification approach \cite{hayes} that was introduced to classify the impact of different levels of clinically significant electromagnetic interference with pacemakers. The monetary, quality of life and privacy impacts are measured using the Fair Information Practice Principles 199 (FIPPS 199) from the  National Institute of Standards and Technology (NIST). The  FIPPS 199 is a standard for assessing the security of information systems. The impact is quantified according to a four-level scale  described in Table~\ref{tab:ImpactLevels} and discussed in more detail in   Section~\ref{sectionImpact}.
\item[Step 2. Scenario-based risk analysis] Here, we estimate the probability of occurrence of various threats.  We start by identifying attack vectors,   i.e.~exploitable vulnerabilities, associated with CIED. We found those attacks vectors on the literature \cite{EcosystemER,halperin2008pacemakers,marin2016security,hei2010defending}, the ICS-CERT advisories \cite{EPDadvisory,ICDadvisory,HMDadvisory}, the National Vulnerability Database (NVD) maintained by the NIST, and the Common Vulnerabilities and Exposure (CVE) database maintained by the Mitre Corporation \cite{CVE1,CVE2,CVE3,CVE5,CVE6,CVE7,CVE13,CVE14} .    Next, we describe how these attack vectors can be strung together into a series of actions,   i.e.~attack scenarios, that lead to the achievement of the attack goals   (determined in Step~1). Once this is done, we calculate for each threat,   i.e.~each (actor, scenario) pair, its probability of occurrence according to the formula
  \begin{equation}
    P = c+o+m  \label{eq:1}
  \end{equation} 
  where $c$, $o$, and $m$ represent, respectively, an assessment of the actor's   \emph{capacity}, \emph{opportunity} and \emph{motivation} to conduct the attack scenario described.  More precisely, capacity takes into consideration the technical complexity of the attack scenario and the technical and material resources available to the actors to carry it out.  The opportunity represents the actor's chances of having physical or network access to the target and being there at the right time to exploit an attack vector and conduct subsequent scenario actions.  Finally, motivation captures the inherent likelihood that the actor will put the resources in place and attempt to conduct the attack scenario given what he stands to gain from successful   accomplishment of the attack goal.\\

\item[Step 3. Combined risk assessment] In this last step, we calculate the overall risk associated with each attack scenario based on the most likely actor.
  \begin{equation}
    R = I*P_{\rm MAX}  \label{eq:2}
  \end{equation}
  Where $I$ is the impact calculated from Step~1, and $P_{\rm MAX}$ is the   maximum actor probability for each attack scenario, as determined in Step~2. %Finally, we calculate the overall risk associated with each attack vector by aggregating the risk over all scenarios that involve that attack vector. 
\end{description}

\section{Actor-based analysis}
\subsection{Potential actors}\label{actors}
\noindent The ICS-CERT has characterized a \emph{cyber threat source} as ``\emph{persons who attempt unauthorized access to a control system device and/or network using a data communications pathway}'' \cite{cyberthreatsourcedescriptions}. It further classifies these threat source into four groups (A1 through A4):
 
\begin{description}
\item[A1. Cybercriminals groups] This includes traditional cybercriminals groups that use compromised computer systems to commit identity theft and online fraud of various kinds, mostly for monetary gain.
\item[A2. Industrial spies] Organizations that use computer tools to illegally acquire intellectual property, know-how, trade, and commercial secrets, or other kinds of corporate confidential information. This kind of espionage occurs between competing corporations, for economic reasons.
\item[A3. Foreign Intelligence Agencies] Foreign state-based organizations that use computer tools to acquire sensitive information on opposing states,   corporations or individuals, or otherwise influence their actions.
\item[A4. Terrorist groups] Organizations seeking to create public disorder or sow national terror, by committing destructive violent acts.
\end{description}

\noindent While this taxonomy of cyber threat sources was introduced for traditional threats to IT infrastructure, we nonetheless proposed to use them in the context of cyber threats against the CIED ecosystem.  To that effect, and considering the likely objectives \cite{bsi} and motivations \cite{cyberthreatsourcedescriptions} of these actors, we maintained the six kinds of attack goals (G1 through G6) described herein.\\

\subsection{Attack goals}\label{sectionGoals}
\subsubsection*{G1. Access patient sensitive data }
\noindent CIED ecosystem devices are an attractive target because they constitute a rich source of information. Beyond medical data, they store other types of information such as email addresses, residence addresses, telephone numbers, social security numbers, etc., attractive to many actors.  On the one hand, intelligence services (A3) and terrorist groups (A4) would be interested in having this information because it would allow them to attain their ultimate goal (surveillance, assassination, etc.). On the other hand, cybercriminal groups (A1) would be interested to leverage this information to obtain monetary gain since the medical data of individuals is highly valued in the black market \cite{ValuePatientsData2,ValuePatientsData3, ValuePatientsData4, ValuePatientsData5}.  Their clients could be for example insurance companies (medical or automotive) that may use this information to assess the cost of insurance premiums or simply refuse coverage.\\

\subsubsection*{G2. Gain knowledge of device operation and software}
\noindent There is significant competition between medical-device manufacturers because of its high-profit margins and high barriers to entry in the market \cite{MedicalDevicesData,lind2017understanding}. Accordingly, CIED ecosystem devices could be a target for industrial spies (A2) aiming to obtain intellectual property on device design, software and other kinds of engineering details.  Subsequently, such information could be sold to competing medical-device manufacturers or possibly to counterfeit medical devices manufacturers in less regulated countries (similarly to the production of counterfeit or generic pharmaceutical products).  Furthermore, this information is also valuable for criminal groups (A1), intelligence services (A3) and terrorist groups (A4) because it allows them to undertake attacks by maliciously exploiting the device characteristics or operating mode.\\

\subsubsection*{G3. Induce medical staff to make errors} \noindent Health is one of the  main factors of concern for individuals.  Hospitals and their personnel are highly valued in society because individuals trust them \cite{halligan2008importance,pilgrim2010examining,van2006public,de2018truth}.  Some attackers may be interested in damaging the reputation of health centers or professionals to sow distrust and fear in the society.  These could include foreign intelligence services (A3), terrorist groups (A4), or even cybercriminal groups (A1).  Apart from sowing fear, said actors could be interested in harming a particular, targeted person.  Thus, inducing medical staff to make errors not only would they be achieving their goal, but they would also be evading the responsibilities of their actions by making their interference less detectable.\\

\subsubsection*{G4. Disrupt or lower quality of patient follow-up}
\noindent Cybercriminal groups (A1) could be attracted by those kinds of attacks to realize extortion, industrial or corporate sabotage.  The first objective would be to disrupt or even interrupt a healthcare provider's service to demand money to restore it.  The second would be to interfere with a targeted CIED manufacturer in order to make believe that their CIED equipment is defective, thus damaging the company's sales revenue and reputation.  The third would be to damage the reputation of a targeted health center or professional. For example, disrupting the quality of a target center's patient follow-up could decrease public and government trust in the institution, and lowering its chances of getting adequate revenue and allocation of public resources.  Intelligence services (A3) or terrorists (A4) could be motivated to conduct such attacks to harm the population of an opposing country.\\

\subsubsection*{G5. Alter device behaviour to endanger patient}
\noindent This constitutes the most potentially worrisome outcome of the cyber attack against the CIED ecosystem.  Indeed, by changing the device settings so that it has an unexpected or dangerous behavior, actors could seriously endanger a patient's life.  It is conceivable that foreign intelligence services (A3) and terrorist groups (A4) targeting particular high-value or highly-visible individuals might be motivated to use this kind of attack for assassinations or as a form or extortion or ransom.\\

\subsubsection*{G6. Alter device behaviour to decrease quality of life}
\noindent For the same reasons described above, intelligence services (A3) and terrorist groups (A4) could be motivated to use similar methods to accomplish non-lethal disruptive effects on patients by forcing them to repeatedly visit the clinic due to device malfunction, generate false alarms, or otherwise tampering with device configuration.  Beyond serious harm, such disruptions could be used to mine the confidence of the population on health providers, device manufacturers, or create panic and terror (A3 and A4).  The possibility should also be considered that cybercriminals (A1) migrate from traditional forms of IT-based extortion, such as file-encrypting ransomware, to medical device-based extortion, e.g.~by locking out access by health practitioners to a patient's CIED and demanding a ransom to restore it.
\\
In summary, the vulnerability of the CIED ecosystem to cyber attacks is a matter of concern not only for patients but also for other groups such as health practitioners, medical device manufacturers and government in general.

\subsection{Impact of attack goals}\label{sectionImpact}
% While we have seen in Section~\ref{actors} that 

% we have seen that actors attack for different
% reasons. Some to earn money, others to get information, and there are also
% others who do it for political reasons. In the section \ref{sectionGoals} we
% note that different groups of actors may have the same attack goal. Therefore,
% an attack can have several effects i.e. impacts.
%%%%%%%%%%%%%%%%%%%%%%%%%%%%%%%%grammar%%%%%%%%%%%%%%%%%%%%%%%%%%%%%%%%%
\noindent Independently of the various actors goals and motivations, these attacks will have an impact on the victim, whether the patients themselves or those other groups affected.  In order to account for the various types of consequences that these attacks could have on them, we measure impact according to four separate aspects: health (H), monetary (M), privacy (P) and quality of life (QL).  We chose these four factors because affecting them negatively align precisely with the attack goals we have previously discussed in Section~\ref{sectionGoals}. Furthermore, by separating our analysis for these factors, we aim to support different agendas and objectives of those organizations potentially interested in this kind of risk assessment, e.g.~health regulation agencies, device manufacturers, health practitioners, etc.  The impact scale ranges from 1 to 4, with 4 being the highest impact level (most severe). The description of the impact levels can be found in Table \ref{tab:ImpactLevels} and the summary of the analysis is presented in the Table \ref{tab:impactResults}. The explanation of the impact analysis by attack goal follows.

\begin{table}[]
       \begin{tabular}{l c c c c} \toprule 
         \textbf{Attacks goal} & \textbf{H} & \textbf{M} &\textbf{QL} &\textbf{P}  \\ \midrule   
        G1 Access patient sensitive data  & - & 1  & - & 2  \\
        G2 Gain knowledge of device operation and software & - & 4  & - & -\\ 
        G3 Induce medical staff to make  errors  & 4 & 3  & 1 & - \\
        G4 Disrupt or lower quality of patient follow-up &2 & 3  & 1 & - \\
        G5 Alter device behaviour to endanger patient & 4 & 3  & - & -\\ 
        G6 Alter device behaviour to decrease quality of life  & - & 2 & 2 & - \\ \hline
    \end{tabular}
\caption{Impact results by attack goal}
\label{tab:impactResults}
\end{table}

\begin{description}
\item[G1] (P) While confidential, the information disclosed would not have a severe consequences (except maybe in terms of insurability) and is likely to   exist in other or be otherwise available to actors through other sources or   other more traditional forms of cyber attacks no related to CIED.  (M) The disclosure of this information may be grounds for legal action against the   hospital and the manufacturer.
\item[G2] (M) The medical device industry is very profitable, and competition   between manufacturers is fierce.  Losses due to intellectual property theft   could reach tens of millions of dollars.
\item[G3] (H) We consider the worst case scenario: the dependent patient (i.e.~one that cannot survive without the device) for whom the doctor does not   make the appropriate diagnosis potentially leading to loss of life.  (M) The doctor and hospital could face severe penalties. (QL) The patient's quality of life would be affected if G3 is achieved.
\item[G4] (H) An interruption in the patient's follow-up could have harmful   effects on the patient's health, for example if an arrhythmic even occured and   there is delay in initiating treatment. (M) For certain health services, time is   money: the interruption of a service for a long period of time can produce losses of millions of dollars for the health organization. (QL) The patient's   quality of life would be affected because of the long waits at the hospital or   the increase in the number of hospital visits.
\item[G5] (H) Worst case scenario, death of dependent patients. (M) In the event   of a legal action, the company could face significant economic penalties.   Moreover, the manufacturer could lose market share or have its devices removed   from the market by regulators.
\item[G6] (M) The equipment could be removed from the market, causing economic   losses to the company. (QL) The patient would feel a temporary discomfort.
\end{description}

\section{Scenario-based risk analysis}
\subsection{Vulnerabilities}\label{Vulnerabilities}
\noindent We now inventory the vulnerabilities ($V_{i}$) affecting the CIED ecosystem. We have harvested this information from several sources, including ICS-CERT advisories, the NVD maintained by the NIST, the CVE database maintained by the Mitre Corporation and previous research in this area \cite{EcosystemER,halperin2008pacemakers,marin2016security,hei2010defending}. We separated the vulnerabilities in three groups, depending on what devices they affect, with some of them applicable to more than one type of device (i.e.~$V_{9}$, $V_{10}$).  We have inventoried 15 vulnerabilities, enumerated in Table~\ref{tab:list} \footnote{In this dissertation, we maintained the original names of the vulnerabilities, i.e., the technical names with which they appear on the source where we extracted them.}, and explained in detail in the following paragraphs.

\begin{table}[]
\centering
  \begin{tabular}{l l l}\\ \toprule
    & \textbf{Vulnerability description} \\ \midrule
    & \emph{CIED}       \\
    $V_1$      & Weak authentication algorithms\\ 
    $V_2$      & Boundless telemetry session duration\\
    $V_3$  & Unencrypted data storage and transmission\\ 
    $V_4$      & Lack of command whitelisting     techniques\\
    & \\\midrule
    & \emph{Programmer}   \\          
    $V_5$      & Unencrypted hardcoded authentication credentials\\   
    $V_6$      & Software directory path traversal\\ 
    $V_7$      & Improper restriction of communication channel \\  
    $V_8$      & Unprotected removable media/hard-drives\\
    $V_9$      & Unprotected USB serial port connections\\ 
    $V_{10}$   & Exploiting embedded debugging interfaces (JTAG and UART)\\ 
    &\\\midrule
    & \emph{Monitor} \\

    $V_9$      & Unprotected USB serial port connections\\ 
    $V_{10}$   & Exploiting embedded debugging interfaces (JTAG and UART)\\ 
    $V_{11}$   & OS hardcoded authentication credentials\\
    $V_{12}$   & Exposed dangerous methods or functions\\     
    $V_{13}$      & Server hardcoded authentication credentials\\     
    $V_{14}$      & Hardcoded server parameters \\
    $V_{15}$      & Exploiting remote firmware update\\
    \bottomrule
\end{tabular}

    \caption{List of vulnerabilities}
    \label{tab:list}
\end{table}

\begin{list}{$V_{\arabic{qcounter}}$:~}{\usecounter{qcounter}}
\item \textbf{Weak authentication algorithms} \\ Certain CIED use   Time-based One-time Password (TOP) for authentication.  The external devices   authenticate to the CIED by computing a password from the current time and a   shared secret, i.e.~a secret cryptographic key shared between the CIED and   both the external programmer and the home-monitoring device, for certain CIED the secret key is their serial or model number. TOP   authentication algorithms are vulnerable to identity theft attacks since an   adversary who steals the secret key can generate valid passwords every time he  wants to establish a telemetry session with the device   \cite{ruxconbreakpoint2012,BarnabyJackBlog,ICDadvisory,CVE1,halperin2008pacemakers}.
\item \textbf{Boundless telemetry session duration} \\ The number of RF wake-up   commands that a CIED can receive per session is not limited, i.e.~an attacker   can maintain a telemetry session indefinitely active by regularly sending the   aforementioned commands to prematurely reduce the CIED's lifetime   \cite{marin2016security,hei2010defending, ICDadvisory,CVE2}.
\item \textbf{Unencrypted data storage and transmission} \\
  Certain CIED models store and transmit patient information without encrypting   it.  Thus, a nearby attacker may intercept the data exchanged between the CIED   and the programmer or even gain access to the sensitive data stored on the   device by sending an unauthorized RF command   \cite{halperin2008pacemakers,ruxconbreakpoint2012,ICDadvisory,CVE3}.
\item \textbf{Lack of command whitelisting techniques}  \\
  Command whitelisting is a computer protection method based on software   restriction policy rules.  This technique blocks by default the execution of   all the programs contained in the device so that only programs that are the   subject of a policy rule can be executed.  In the case of CIED there are no   policy rules prohibiting the execution of programming commands from devices   other than external programmers.  Consequently, an adversary could send a   programming command to the CIED by means of commercial available equipment   such as a commercially-available SDR   \cite{halperin2008pacemakers,marin2016security,ruxconbreakpoint2012}.
\item \textbf{Unencrypted hardcoded authentication credentials}\\ %(CVE-2018-5446)
  The product username and password are stored in a recoverable format, i.e.~without being previously encrypted \cite{EPDadvisory,CVE5}.
\item \textbf{Software directory path traversal} \\
  It has been shown that the software of certain devices contain directory path   traversal vulnerabilities, i.e.~a kind of software implementation   vulnerability that permits the access to directories other than those   permitted by design. Thus, an adversary will be able to exploit these   weaknesses in order to read the external programmer's file system   \cite{EPDadvisory,CVE6}.
\item \textbf{Improper Restriction of Communication Channel}\\
  Downloading software updates is done by means of a Virtual Private Network   (VPN) established between the programmer and its software update provider.   While the use of VPN is a recognized good practice to secure communications   between two parties, it has been unveiled that certain external programmers   models do not verify that they are still connected to the VPN before the   update operation is accomplished.  Thus, an adversary could leverage the   device's local network access features to interfere with the communication   between the programmer and its software update provider   \cite{EcosystemER,EPDadvisory,CVE7}.
\item \textbf{Exploiting embedded debugging interfaces (JTAG and UART)}\\
  Embedded debugging interfaces are connection ports present in a device's   printed circuits. Manufacturers use them to perform functional testing and   redesign of devices after manufacturing. For example, JTAG is a master/server   interface used to verify a circuit, test device logic and perform functional   redesign when needed.  It can be used to read and modify the memory and the   registers as well as to read the device's firmware.  The UART interface   provides a serial communication between the device's embedded systems and an   external PC, i.e.~a bidirectional interface used to send and receive data   asynchronously.  Since these interfaces allow direct access to the device   memory and firmware, unprotected access to those interfaces constitutes an entry point for attacks against the CIED \cite{EcosystemER}.  Home monitoring devices   also have this vulnerability.
\item \textbf{Unprotected USB serial port connections}\\
  Certain devices have USB port connections.  They are frequently used by   medical staff to store the information on a USB stick in order to transfer it   to other systems, e.g.~reporting software. If the USB port connection is not   blocked with a password or another authentication mechanism, an attacker could   connect to it and access data on the device and potentially take control of   it\cite{EcosystemER}.
\item \textbf{Unprotected removable media/hard-drives}\\
  When they are in the attacker's hands, the media/hard drives become an entry point of attacks since they can be used to extract information from a device's file   system \cite{EcosystemER}.
\item \textbf{OS Hard coded authentication credentials} \\
  In certain products, authentication credentials to the operating system (OS)   are hard-coded on the device. That means that an adversary with physical access to the device's integrated circuit can access the OS by connecting to   the debug port and authenticate with the hard-coded password   \cite{EcosystemER,HMDadvisory,CVE13}.
\item \textbf{Exposed dangerous methods or functions} \\
  Home monitors contain debug code to test their communication interfaces with   both the CIED or the external system (databases, servers) of the cloud-based   application used by the physicians.  Thus, by leveraging this vulnerability an   adversary with physical access to the monitor can maliciously exploit the   debug code to accomplish a set of attacks, for example, read or write the   device's memory content, interrupt the data sending to the cloud-based   systems, enable bidirectional communication with CIED   \cite{HMDadvisory,CVE14}.
\item \textbf{Server hardcoded authentication credentials}\\
  The credentials that home monitors use to authenticate to the cloud-based   systems supporting the patient's remote follow-up service are hard-coded on   certain devices. Thus, an attacker with physical access to the monitor can   leverage these vulnerabilities to access the database in order to read or   tamper with the patient's medical data \cite{EcosystemER}.
\item \textbf{Server hardcoded parameters}\\
  In certain home monitors the IP address of the authentication servers are   hard-coded. An adversary could use this information to conduct a DoS attack to   make the server temporarily unavailable by sending several web requests to   this IP address \cite{EcosystemER}.
\item \textbf{Exploiting remote firmware update}\\
  Firmware updates for home monitors are triggered remotely. Indeed, when the time comes to update the device's firmware, the manufacturer sends the new   version to the monitor through the cloud.  This method is advantageous from   the patient's point of view since it avoids an additional trip to the   hospital. However, it constitutes at the same time an attack vector because the home-monitoring device does not verify the identity of the system distributing the firmware. An attacker could take advantage of this lack of verification by achieving a man-in-the middle attack with the purpose of sending a counterfeit firmware to the device \cite{EcosystemER}.
\end{list}

\subsection{Attack scenarios}\label{AttackScenario}
\noindent Once we have identified who the actors are and what they are trying to achieve (attack goals), we are now interested in the strategy that it is going to be used by them, i.e.~how will they exploit the vulnerabilities of the CIED ecosystem to achieve their goals? Thus, as illustrated in Table~\ref{tab:scenario}, an attack goal can be achieved through different scenarios. As defined in Section~\ref{sectionDefinition} an attack scenario is the sequence of events that must occur for the attack to take place.\\

\noindent It can be noticed that the same scenario can serve to achieve different attack goals. Since a threat is a pair (actor, scenario) and the actors can vary from one attack goal to the next, we carried out the scenario-based risk analysis by attack goals. The explanation of the scenarios of each attack goal follows.  For a more extensive description of the sequence of events leading to the achievement of the attack scenarios refer to~\ref{AppendixB}.

\begin{table*}[]
    \centering
     \resizebox{\linewidth}{!}{
    \begin{tabular}{l l l l}\rowcolor{white} \toprule
\textbf{Attack }&\textbf{Scenario}&\textbf{Scenario description}& \textbf{Method}\\
\textbf{goal}&&&\\ \rowcolor{yellow!5} \midrule
&&&\\\rowcolor{yellow!25}
$G1$&$S_1$&CIED-Monitor communication interception& Intercepting RF signals with an SDR\\ \rowcolor{yellow!25}
&$S_2$&CIED-Programmer communication interception& Intercepting RF signals with an SDR\\ \rowcolor{yellow!25}
&$S_3$&Extraction of health data stored into the monitor& Connecting to the debbugging ports\\ \rowcolor{yellow!25}
&$S_4$&Insertion of a backdoor (malware) into the monitor& Performing MITM attack during a \\ \rowcolor{yellow!25}
&&& firmware update session\\ \rowcolor{yellow!10}  
&&&\\ \rowcolor{yellow!25}
$G2$&$S_{4}$&Insertion of a backdoor (malware) into the monitor& Performing a MITM attack during a\\  \rowcolor{yellow!25}
&&&firmware update session\\  \rowcolor{yellow!25} 
&$S_5$&Extraction of the programmer's system data& Sending a malicious http request \\ \rowcolor{yellow!25}
&&from the device's SW deployment network server& to the server \\ \rowcolor{yellow!25}
&$S_6$&Extraction of the programmer's system data & Accessing the device through an  \\ \rowcolor{yellow!25}
&&& update session communication channel \\ \rowcolor{yellow!25}

&$S_7$&Reading/extraction of the monitor file system & Accessing the device USB port \\ \rowcolor{yellow!25}
&$S_{8}$&Reading/extraction of the programmer file system & Accessing he device USB port \\ \rowcolor{yellow!25}
&$S_{9}$&Reading/extraction of the programmer system data& Removing the media device hard drive \\ \rowcolor{yellow!25}
&$S_{10}$&Reading/extraction of the monitor OS information & Connecting to the debugging ports \\ \rowcolor{yellow!10}

&&&\\ \rowcolor{yellow!25}

$G3$&$S_{11}$&  Insertion of a malware that produce programmer &Performing a  MITM attack during an\\ \rowcolor{yellow!25}
&& reading errors& update session\\ \rowcolor{yellow!25}
&$S_{12}$&Introduction of calibration errors into the CIED & Sending  RF  commands \\ \rowcolor{yellow!25}
&& microprocessor (through malware insertion  or sending  & with an SDR \\ \rowcolor{yellow!25}
&& inappropriate commands)&\\\rowcolor{yellow!25}
&$S_{13}$&Insertion of a malware that produce programmer& Using the device USB port \\ \rowcolor{yellow!25}
&&reading errors&\\  \rowcolor{yellow!10} 

&&&\\ \rowcolor{yellow!25}
$G4$ &$S_{4}$&Insertion of a ransomware (malware) into the monitor & Performing a MITM attack during a \\ \rowcolor{yellow!25}
&& &firmware update session\\ \rowcolor{yellow!25} 

&$S_{11}$&Insertion of a ransomware (malware) into the programmer & Performing a MITM attack during an \\ \rowcolor{yellow!25}
&&& update session\\ \rowcolor{yellow!25}

&$S_{12}$& Insertion of a ransomware (malware) into the CIED& Sending RF commands with an SDR \\ \rowcolor{yellow!25}

&$S_{14}$&Maintain CIED telemetry session open indefinitely& Sending RF commands with an SDR \\ \rowcolor{yellow!25}

&$S_{15}$&Modify/erase the contents of the monitor memory& Connecting to the debugging ports \\ \rowcolor{yellow!10}

&&&\\ \rowcolor{yellow!25}

$G5$&$S_{11}$&Insertion of a malware that ignores programmer  & Performing a MITM  attack during an\\ \rowcolor{yellow!25}
&&  therapy settings  &update session\\ \rowcolor{yellow!25}
&$S_{11}$&Insertion of a malware that makes programmer & Performing a MITM attack during \\  \rowcolor{yellow!25}
&&apply a predefined dangerous treatment& update session\\\rowcolor{yellow!25}

&$S_{11}$&Insertion of a backdoor (malware) into the programmer  & Performing a MITM attack during a \\ \rowcolor{yellow!25}
&&&session update\\ \rowcolor{yellow!25} 

&$S_{12}$&Modification of the CIED section of RAM containing & Sending RF unauthorized commands \\ \rowcolor{yellow!25}
&& the therapy code to be applied to the patient&with an SDR \\ \rowcolor{yellow!10}

&&&\\ \rowcolor{yellow!25}
$G6$&$S_{10}$&Disable the periodic data transmission from the monitor &Connecting to the debugging ports\\ \rowcolor{yellow!25}
&$S_{11}$&  Insertion of a malware that produce programmer's & Performing a  MITM attack during an\\ \rowcolor{yellow!25}
&& reading errors& update session\\ \rowcolor{yellow!25}
&$S_{14}$&Maintain a CIED's telemetry session indefinitely open& Sending RF commands with an SDR \\ \rowcolor{yellow!5}

&&&\\\bottomrule 
    \end{tabular}
}
     \caption{Attack scenarios}
    \label{tab:scenario}
\end{table*}

\subsubsection*{G1 Access patient sensitive data}
\noindent There are three ways to acquire patients medical data: performing a radio attack  ($S_{1}$, $S_{2}$) on the incoming RF communication between the CIED and the external devices (monitor, programmer), getting unauthorized physical access to the monitor contents ($S_{3}$) or performing a network attack on the monitor ($S_{4}$).\\

\noindent Executing the radio attacks described in Scenarios $S_{1}$ and $S_{2}$
requires the actor to have specialized materials and software, namely
an SDR, an antenna and a radio signal processing software
(e.g.~GNURadio, HackRF, etc.). Once this requirement has been met, the
actor must go either to the patient's home ($S_{1}$) or to the
hospital($S_{2}$), place himself at a distance relatively close to the
CIED, configure its antenna in reception mode, tune it to the
transmission frequency of the CIED then, record the signals
emitted by the latter and read the patient's medical data by
exploiting the CIED unencrypted data storage and transmission
vulnerability ($V_{3}$).\\

\noindent The physical attack of  Scenario $S_{3}$ also requires the actor to have specialized equipment. An in-debugger-circuit, a debugger IDLE and a pirate bus (or an F to F jumper wire) are needed. Since the monitor is the targeted device, the actor must go to the patient home, then connect to the device's debugging interfaces employing the pirate bus (or the F to F jumper wire). After that, he must use the in-debugger-circuit along with the debugger IDLE to access the monitor's memory content\footnote{This attack scenario will be used especially when the ultimate goal of the attacker would be to subsequently attack the server to obtain the medical data of several patients.}.  Consequently, the actor must exploit the following three vulnerabilities of the monitor: exploiting debugging interfaces ($V_{10}$), server hard-coded authentication credentials ($V_{13}$) and, hard-coded server parameters ($V_{14}$).\\ 

\noindent The monitor is once again the target device in Scenario $S_{4}$.  Here,  the network attack proposed relies on installing a backdoor on the device. In this case, the actor must know beforehand the day when an update will take place. Once done, he must approach the patient's home then, access the patient's private network, and achieve a man-in-the-middle attack exploiting the monitor's remote firmware update session ($V_{15}$). At that point, the actor must swap the updated firmware for a backdoor. Thus, he will be able to access the target at any later time employing the backdoor.\\
%The backdoor access mode is a  hacking technique consisting on bypass the authentication mechanism of a computer system. Hackers use backdoors to gain access to systems or to their data, and sometimes to install malware (malicious software) 

%(MITM) attack\That is a hacking technique to intercept and modify the information transmitted in the communication channel established between two parties. 
\subsubsection*{G2 Gain knowledge of device operation and software}
\noindent G2 can be achieved by performing network attacks on the external devices ($S_{4}$, $S_{6}$, $S_{7}$, or $S_{8}$), launching a web attack on the programmer software deployment network server  ($S_{5}$), or getting  unauthorized physical access to the external devices ($S_{9}$, $S_{10}$). In the last case we will talk about a physical attack on the external devices.\\

\noindent For the network attacks of Scenarios  $S_{4}$, $S_{6}$, $S_{7}$, and $S_{8}$, the actor must either go to the patient's home ($S_{4}$, $S_{7}$) or the hospital ($S_{6}$, $S_{8}$). Note that for  Scenarios $S_{4}$ and $S_{7}$, this must occur the day of an update of the monitor and the programmer respectively. Once on the crime scene, the actor should access either the targeted device network  ($S_{7}$, $S_{8}$) or the communication channel established between the communicating parties  ($S_{4}$, $S_{6}$). In the last case, the communicating parties are the external device and the web server of the entity in charge of the updates. Thus, once in the external device network the actor should either connect himself to the USB port and acquire the file system  ($S_{7}$, $S_{8}$) or have direct access to the devices and therefore to the data  ($S_{4}$, $S_{6}$).\\

\noindent In  Scenario $S_{5}$ the actor must find the URL from which the programmer update application retrieve files from the server of the software deployment network. Once this is done, he  modifies the URL with commands and web server escape code.  After that, he sends this URL to the web server by means of a web request. Thus, if the attack is successful, the actor will be able to extract the desired files.\\

\noindent Getting an unauthorized physical access to the external devices  ($S_{9}$ and $S_{10}$) is another mean to achieve G2. On Scenario $S_{9}$, the extraction of the programmer hard drive is required. Thus, the actor should go to the hospital and remove it. As far as Scenario $S_{10}$ is concerned  the attack is on the monitor, that is to say that the crime scene is the patient's home. The sequence of events of this scenario is that of  $S_{3}$ except that two events are added, namely 1) connect to the debug port of the operating system and then 2) authenticate using the credentials that will have been previously acquired by performing the same actions as in  $S_{3}$.\\

\subsubsection*{G3 Induce medical staff to make diagnostic errors}

\noindent G3 can be achieved is feasible by achieving three kinds of attacks: a network attack on the programmer ($S_{11}$), a radio attack on the CIED ($S_{12}$) or a  physical attacks on the programmer ($S_{13}$).\\

\noindent The sequence of events for Scenario $S_{11}$ is practically the same as that for Scenario $S_{4}$. What differentiates both scenarios is the target device.  In $S_{4}$, it is the monitor while  in  $S_{11}$, it is the programmer.  Thus the only difference between  $S_{11}$ and  $S_{4}$ stems from the first event, that in the case of $S_{11}$ is happening in the patient's home.\\

\noindent Scenario $S_{12}$ is completely similar to Scenarios $S_{1}$ and $S_{2}$.
The only change is the actor's behavior.  Indeed,  in Scenarios $S_{1}$ and $S_{2}$ he intercepts data;  he is a passive actor.  In Scenario  $S_{12}$, however, he transmits  data, thus he is an active actor. The events in Scenario $S_{12}$  are otherwise practically the same events as in $S_{1}$ and $S_{2}$. We say practically because first, a new event is added. That is the transmission of data. Second, one of the events of $S_{1}$ and $S_{2}$ is modified. In fact we saw for the G1 scenario the actor would have to configure his antenna in reception mode to intercept the data, while in  $S_{12}$ it will have to put in transmission mode.\\

\noindent In Scenario $S_{13}$ a network attack is performed on the programmer. The actor's purpose here is to introduce a calibration error on the device, by inserting a malware through the device's UBS port connection. In order to do so, he goes to the patient's home, accesses the patient's network, scans the network ports in order to find the one that corresponds to the USB connection, then sends the malware by means of the aforementioned port. As it can be noticed, the sequence of events for Scenario  $S_{13}$ is quite similar  Scenario $S_{8}$. The difference between both is the last event which in $S_{8}$ is accessing the device file system while in $S_{13}$ it is sending the malware.\\

\subsubsection*{G4 Disrupt or lower quality of patient follow-up}
\noindent The goals of G4 can be accomplised by performing  a network attack against the external devices ($S_{4}$, $S_{11}$), radio attacks against the CIED ($S_{12}$, $S_{14}$) or physical attack against the monitor.\\

\noindent In the first cases, i.e.~Scenarios $S_{4}$ and $S_{11}$, the purpose
of the attack is to render the data of the external devices
unreadable. To do this, the actor will send a ransomware to the
devices, i.e.~a kind of malware that encrypts the system data. Data
restoration consists of applying the same operation to the encrypted
data with decryption key.  Normally, the malware operator will have generated and kept secret a copy of the  decryption key, that will only be revealed to the victim in exchange for ransom.  The sequence of the events is similar to that of $S_{4}$ and
$S_{11}$ for the G1 and G2 scenarios. The difference lies in the type
of malware used, and this has no effect on the sequence of events.\\

\noindent For the  radio attacks, the sequence of events leading to $S_{12}$ because is similar to that  in G3. What changes between the two attacks goals is the nature of the data transmitted by the actor. In G3, it is a dangerous command, here it is malware. Scenario $S_{14}$ consists off periodically sending wake-up commands to the CIED to maintain open the incoming wireless communication. In order to do that, the actor must obtain an SDR, an antenna and a signal processing software. He must track the victim and replay an RF wake-up command every time the wireless session is about to expire.\\   

\subsubsection*{G5 Alter device behaviour to endanger patient }\label{scenarioG5}
\noindent G5 is achievable by perpetrating network attacks on the programmer
($S_{11}$). These attacks can take several forms as detailed in the
Table~\ref{tab:scenario}. Indeed, the actor can implement these
scenarios to send malicious code that ignores the therapy settings set
by the practitioners, or introduces a calibration error into the
device, or allows him to access the device by means of a
backdoor. Performing a radio attack against the CIED ($S_{12}$) is
another way to accomplish the Goal G5. The actor's purpose here will be
to modify the device's RAM section containing the therapy code to be
applied to the patient. As those scenarios have already been appearing
in previous attacks goals scenarios (G3 and G4), the event sequence
will be the same.\\

\subsubsection*{G6 Alter device behaviour to decrease quality of life} 
\noindent Three kinds of attacks can be carried out in order to achieve attack
Goal G6. The first one, $S_{10}$, consists in perpetrating a physical
attack on the monitor with the purpose of disabling the device's
periodic data transmission. The second one, $S_{11}$, relies on the
execution of a network attack on the programmer. The actor introduces
a calibration error on the device by inserting malware. The third one,
$S_{14}$, is a radio attack on the CIED. The goal will be to maintain
a wireless communication session indefinitely open by sending RF
wake-up commands. The event sequence is similar to that of Goal G5.\\

\subsection{Probabilities of Occurrence}\label{sectionProbability}

\noindent As defined in Section~\ref{sectionDefinition}, the probability of occurrence
($P_{r}$) represents the chance that a given threat (actor-scenario pair)
materializes. In other words, it is the likelihood that an actor achieves an
attack scenario with success. By success we mean the achievement of the attack's
goal or what is the same, the engendering of a specific impact on the victim. We
calculate the probability by threat. That is, for each actor of each
scenario. As explained in the methodology section (Section~\ref{methodology}),
$P_{r}$ is calculated (\ref{eq:1}) as the sum of the three threat attributes:
capacity ($c$), opportunity ($o$) and motivation ($m$). The $c$, $o$, $m$ values
vary from 1 to 4, with 4 corresponding to a higher likelihood.  In the following
paragraphs, we justify the rates assigned to $c$, $o$, $m$ for each threat, with the overall
$P_{r}$ values given in Table~\ref{tab:probability}.

\begin{table}[]
\resizebox{0.5\linewidth}{!}{
    \begin{tabular}{l l l l l l l}\rowcolor{green!3}\toprule
    \textbf{Attack } & \textbf{Scenario} & \textbf{actor} & $c$&$o$&$m$&$P_r$  \\\rowcolor{green!3} 
\textbf{goal} & & &&&& \\ \rowcolor{green!3} \hline
&&&&&&\\ \rowcolor{green!25}
$G1$&$S_1$      & A3  & 3  & 2  & 2 & 7\\\rowcolor{green!25}
&           & A4  & 3  & 1  & 1 & 5\\\rowcolor{green!25}
&$S_2$ & A3  & 3  & 2  & 2 & 7  \\ \rowcolor{green!25}
&           & A4  & 3  & 2  & 1 & 6\\\rowcolor{green!25}
&$S_3$      & A3  & 2  & 2  & 2 & 6\\ \rowcolor{green!25}
&           & A4  & 1  & 1  & 1 & 3\\ \rowcolor{green!25}
&$S_4$      & A3  & 3  & 2  & 2 & 7\\ \rowcolor{green!25}
&           & A4  & 3  & 1  & 1 & 5\\ \rowcolor{green!3}
&&&&&&\\ \rowcolor{green!25}
$G2$&$S_{4}$   & A1  & 4  & 1  & 2  & 7 \\ \rowcolor{green!25}
&           & A2  & 3  & 2  & 4  & 9\\ \rowcolor{green!25}
&           & A3  & 3  & 2  & 3  & 8\\ \rowcolor{green!25}
&           & A4  & 3  & 1  & 3  & 7\\ \rowcolor{green!25}
&$S_5$      & A1  & 4  & 3  & 2 & 9  \\ \rowcolor{green!25}
&           & A2  & 4  & 3  & 4 & 11\\ \rowcolor{green!25}
&           & A3  & 4  & 3  & 3 & 10\\ \rowcolor{green!25}
&           & A4  & 3  & 3  & 3 & 9\\ \rowcolor{green!25}
&$S_6$      & A1  & 4  & 1  & 2 & 7\\ \rowcolor{green!25}
&           & A2  & 3  & 2  & 4 & 9\\ \rowcolor{green!25}
&           & A3  & 3  & 2  & 3 & 8\\ \rowcolor{green!25}
&           & A4  & 3  & 1  & 3 & 7\\ \rowcolor{green!25}
&$S_7$      & A1  & 4  & 2  & 2 & 8\\ \rowcolor{green!25}
&           & A2  & 3  & 3  & 4  & 10\\ \rowcolor{green!25}
&           & A3  & 3  & 3  & 3  & 9\\ \rowcolor{green!25}
&           & A4  & 3  & 2  & 3  & 8\\ \rowcolor{green!25}
&$S_{8}$ & A1  & 4  & 3  & 2  & 9   \\ \rowcolor{green!25}
&           & A2  & 3  & 3  & 4  & 10\\ \rowcolor{green!25}
&           & A3  & 3  & 3  & 3  & 9\\ \rowcolor{green!25}
&           & A4  & 3  & 3  & 3  & 9\\ \rowcolor{green!25}
&$S_{9}$   & A1  & 4  & 1  & 1  & 6\\ \rowcolor{green!25}
&           & A2  & 4  & 2  & 1  & 7\\ \rowcolor{green!25}
&           & A3  & 4  & 2  & 1  & 7\\ \rowcolor{green!25}
&           & A4  & 4  & 1  & 1  & 6\\ \rowcolor{green!25}
&$S_{10}$   & A1  & 1  & 1  & 1  & 3\\ \rowcolor{green!25}
&           & A2  & 2  & 2  & 1  & 5\\ \rowcolor{green!25}
&           & A3  & 2  & 2  & 1  & 5\\ \rowcolor{green!25}
&           & A4  & 1  & 1  & 1  & 3\\ \rowcolor{green!25}
&           &   &   &   &   & \\ \rowcolor{green!25}
&           &   &   &   &   & \\ \rowcolor{green!25}
&           &   &   &   &   & \\ \rowcolor{green!25}
&           &   &   &   &   & \\ \rowcolor{green!25}
&           &   &   &   &   & \\ \rowcolor{green!25}
&           &   &   &   &   & \\ \rowcolor{green!25}
&           &   &   &   &   & \\ \rowcolor{green!3}
&&&&&&\\ 
\hline
\end{tabular}
}
%\end{table}
%\begin{table}[ht!]
\resizebox{0.5\linewidth}{!}{
    \begin{tabular}{l l l l l l l} \rowcolor{green!3}  \toprule
   \textbf{Attack } & \textbf{Scenario} & \textbf{actor} & $c$&$o$&$m$&$P_r$  \\\rowcolor{green!3} 
\textbf{goal} & & & &&&\\ \rowcolor{green!3}\hline
&&&&&&\\ \rowcolor{green!25}
$G3$&$S_{11}$   & A1  & 4  & 1  & 1  & 6\\ \rowcolor{green!25} 
&           & A3  & 3  & 2  & 3  & 8\\ \rowcolor{green!25}
&           & A4  & 3  & 1  & 3  & 7\\  \rowcolor{green!25}
&$S_{12}$   & A1  & 1  & 1  & 1  & 3\\ \rowcolor{green!25}
&           & A3   & 2  & 2  & 3  & 7\\ \rowcolor{green!25}
&           & A4  & 2  & 1  & 3  & 6\\ \rowcolor{green!25}
&$S_{13}$   & A1  & 4  & 3  & 1  & 8\\ \rowcolor{green!25}
&           & A3  & 3  & 3  & 3  & 9\\ \rowcolor{green!25}
&           & A4  & 3  & 3  & 3  & 9\\ \rowcolor{green!3} 
&&&&&&\\ \rowcolor{green!25}
$G4$&$S_{4}$   & A1  & 3  & 1  & 1  & 5\\ \rowcolor{green!25}
&           & A3  & 2  & 2  & 3  & 7\\\rowcolor{green!25}
&           & A4  & 2  & 1  & 2  & 5\\ \rowcolor{green!25}
&$S_{11}$   & A1  & 3  & 1  & 1  & 5\\ \rowcolor{green!25}
&           & A3  & 2  & 2  & 3  & 7\\\rowcolor{green!25}
&           & A4  & 2  & 1  & 2  & 5\\  \rowcolor{green!25}
&$S_{12}$   & A1  & 3  & 1  & 1  & 5\\ \rowcolor{green!25}
&           & A3  & 2  & 2  & 3  & 7\\\rowcolor{green!25}
&           & A4  & 2  & 1  & 2  & 5\\  \rowcolor{green!25}
&$S_{14}$   & A1  & 3  & 1  & 1  & 5\\ \rowcolor{green!25}
&           & A3  & 3  & 2  & 2  & 7\\\rowcolor{green!25}
&           & A4  & 3  & 1  & 3  & 6\\ \rowcolor{green!25}
&$S_{15}$   & A1  & 1  & 1  & 1  & 3\\ \rowcolor{green!25}
&           & A3  & 3  & 2  & 3  & 8\\\rowcolor{green!25}
&           & A4  & 2  & 1  & 2  & 5\\  \rowcolor{green!3} 
&&&&&&\\ \rowcolor{green!25}
$G5$&$S_{11(a)}$   & A3  & 3  & 2  & 2  & 7\\\rowcolor{green!25}
&           & A4  & 3  & 1  & 3  & 6\\  \rowcolor{green!25}
&$S_{11(b)}$   & A3  & 2  & 2  & 2  & 6\\\rowcolor{green!25}
&           & A4  & 1  & 1  & 3  & 5\\    \rowcolor{green!25}
&$S_{11(c)}$   & A3  & 3  & 2  & 2  & 7\\ \rowcolor{green!25}
&           & A4  & 3  & 1  & 3  & 7\\ \rowcolor{green!25} 
&$S_{12}$   & A3  & 2  & 2  & 2  & 6\\ \rowcolor{green!25}
&           & A4  & 2  & 1  & 3  & 6\\    \rowcolor{green!3}
&&&&&&\\ \rowcolor{green!25}
$G6$&$S_{10}$& A1  & 1  & 1  & 1  & 3\\ \rowcolor{green!25}
&           & A3  & 2  & 2  & 3  & 7\\ \rowcolor{green!25}
&           & A4  & 1  & 1  & 3  & 5\\ \rowcolor{green!25}
&$S_{11}$   & A1  & 4  & 1  & 1  & 6\\ \rowcolor{green!25}
&           & A3  & 3  & 2  & 3  & 8\\\rowcolor{green!25}
&           & A4  & 3  & 1  & 3  & 7\\ \rowcolor{green!25}
&$S_{14}$   & A1  & 3  & 1  & 1  & 5\\ \rowcolor{green!25}
&           & A3  & 3  & 2  & 3  & 8\\\rowcolor{green!25}
&           & A4  & 3  & 1  & 3  & 6\\   \rowcolor{green!3}
&&&&&&\\ \hline
\end{tabular}
}
    \caption{Probability of occurrence of identified threats}
    \label{tab:probability}
  \end{table}
  
\subsubsection{Attack goal G1}
\paragraph*{Capacity}
Scenarios $S_{1}$ and $S_{2}$ are accomplished by means of radio attacks.  The
capacit for Actors A3 and A4 are the same ($c=3$) for many reasons: the
knowledge is abundant and accessible to all the actors, the software tools used
to intercept and process RF signals are increasingly simpler to use, thus
reducing the attack's technical difficulty, and the equipment needed to perform
these attacks (SDR and antenna) is not expensive. For Scenario $S_{3}$, even if
the knowledge is accessible to all the actors and the equipment needed to
conduct the attack is not expensive, the attack is technically complex to
achieve. Indeed, it involves the exploitation of two vulnerabilities for which
solid knowledge of computer programming and architecture is required. Normally,
Actor A3 recruits experts with exceptional technical skills and have more human
resources. They have more capacity than Actor A4. Thus, in Scenario $S_{3}$ A3
capacity ($c=2$) is higher than the one of Actor A4 ($c=1$).  Scenario $S_{4}$
is a network attack and thus additional material is not required.  Additionally,
there is nowadays extensive information available and tools to perform the attack in
$S_{4}$ . Thus, capacity for Actors A3 and A4 will be the same ($c=3$) in this scenario.

\paragraph*{Opportunity}
In Scenarios $S_{1}$ and $S_{3}$, the attack takes place in the patient's
home. In these cases Actor A3 ($o=2$) has a better chance than Actor A4 ($o=1$)
since they are specifically trained to infiltrate private sites without being
noticed. In Scenario $S_{2}$, the attack takes place in the hospital during a
patient's medical visit. The latter implies that adversaries only have
approximately two days a year to conduct the attack, coinciding with the number
of times patients go to the doctor.  However, since hospitals are public places,
the actors are less likely to be noticed. Thus, the opportunity score for
Actors A3 and A4 ($o=2$) is the same. In Scenario $S_{4}$ the attacks take
place during a monitor's update session, which takes place only about once a
year. Actor A3 access to this information and opportunity to leverage it is greater ($o=2$) than that of Actor A4
($o=1$).

\paragraph*{Motivation}
Both Actors A3 and A4 benefit from the crime.  They gain access to sensitive
personal information. For A3, this attack objective is in line with the
\emph{raison d'être} of their profession, i.e.~obtaining private information
from individuals. Thus the motivation of Actor A3 ($m=2$) will be higher than that of Actor A4
($m=$1) because for A3 this attack objective is an end in itself while for A4 it is a means
to an end (sow national disorder).

\subsubsection{Attack goal G2}\label{probG2}
\paragraph*{Capacity}
In Scenario $S_{5}$ a web attack is launched. There is information and tools
available online to perform this kind of attack.  Actor type A4 are experts in
the field (web attack).  On the other hand, Actors A2 and A3 are specialists in
the extraction of information from people or systems.  In addition, they often
have specialized human resources.  Thus the capacity of A1, A2 and A3 ($c=4$) is
the same and it is higher than that of A4 ($c=3$). In Scenarios $S_{4}$,
$S_{6}$, $S_{7}$ and $S_{8}$ network attacks are conducted.  Once more,
information and tools are available to achieve these attacks. However, because
they have more know-how than the others on the matter (i.e.~network attacks)
Actor A4's capacity ($c=4$) is higher than that of A1, A2 and A3 ($c=3$). The
attack performed in $S_{9}$ has no major technical complications. It is
necessary to remove a hard disk and then mount it later in another computer
media. Thus, the capacity of all actors will be the same ($c=4$). However, the
achievement of Scenario $S_{10}$ presents a major challenge. On the one hand,
solid technical knowledge of computer programming and architecture is
necessary. In addition, there is no extensive information about how to realize
the exploit in $S_{10}$ . Thus, Actors A2 and A4's capacity ($c=2$) is higher
than that of A1 and A4 ($c=1$) since A2 and A3 normally are experts with
exceptional technical skills and have more human resources.

\paragraph*{Opportunity}
Scenario $S_{5}$ is a web attack where there is no restriction of time and space. So the actors' opportunity will be higher and the same ($o=3$).  Scenarios $S_{6}$ and $S_{4}$ take place during targeted device update sessions, during which there are constraints in terms of time (update session) and space (near the patient's home or hospital).  As far as the time constraint is concerned, Actors A2 and A3 have better possibilities to know when an update session will take place. In terms of space constraint, A2 and A3 have the same opportunities either at the patient's home or in the hospital. However, A1 and A4 will have more chances in the hospital as this is a public place where they can go unnoticed. Thus on the  $S_{6}$ and $S_{4}$ Scenarios, Actors A2 and A3 opportunity is higher ($o=2$) than that of A1 and A4 ($o=1$). For  Scenarios $S_{7}$, $S_{8}$, $S_{9}$ and $S_{10}$ there is no time constraint  but there is still a space constraint.  Scenarios $S_{7}$ and $S_{8}$ require the actor to be near either the patient's home or the hospital in order to access their network, whereas for $S_{9}$ and $S_{10}$ the actor must to have physical access to the targeted devices. Similarly as for  $S_{7}$, since the attack takes place near to patient's home Actors A2 and A3 opportunity ($o=3$) will be higher than the one of Actors A1 and A4 ($o=2$). For $S_{8}$ however, all actors opportunity score  is the same  ($o=3$) since the attack takes place in a public site. In Scenarios $S_{9}$ and $S_{10}$, since the attack requires  physical access to the device Actors A2 and A3 opportunity ($o=2$) is higher tan that of A1 and A4 ($o=1$).

\paragraph*{Motivation}
All actors benefit from the crime. They gain system information. A2 motivation ($m=4$) is the highest since the goal of this attack is the purpose of their profession. Actors  A3 and A4 follow them with the same level of motivation ($m=3$). The motivation of A1 (m=2) is the lowest because obtaining system information is not an end but a means to accomplish their activities.

\subsubsection{Attack goal G3}\label{probG3}

\paragraph*{Capacity}
Attack scenarios $S_{11}$, $S_{12}$ and $S_{13}$ consist in introducing reading
or calibration errors on the CIED's ecosystems devices. To do that knowledge of
the device inner workings and advanced programming skills are required. Since
there is some but not a lot of available information about how programmers and monitors work, in
Scenarios $S_{11}$ and $S_{13}$ the capacity of A1 ($c=4$) will be higher than
that of A3 and A4 ($c=3$). The reason is that A1 are experts in the development of
malicious code.  On the other hand, there is much less  information available about CIED and their architecture. Thus, for  Scenario $S_{12}$, the capacity of the actors will be
the same ($c=2$). This is due to the fact that while A1 are experts in malware
development, A3 and A4 are more likely to obtain the CIED's mode of operation
either by hiring personnel skilled in CIED programming or by using other illegal methods.

\paragraph*{Opportunity}
For Scenarios $S_{11}$ and $S_{12}$ there are constraints in terms of time and space. Scenario $S_{11}$ takes place in the hospital during a session update. Scenario $S_{12}$ must be performed near the patient and during an incoming wireless communication with one of the externals devices. In these scenarios, we apply the same opportunity values that we have applied to the scenarios  $S_{6}$ and $S_{4}$. That is to say that in $S_{11}$ and $S_{12}$, Actor A3's opportunity ($o=2$) is higher than that of Actors A1 and  A4 ($o=1$). On $S_{13}$ there is only a space restriction, and the same reasoning as in Scenario $S_{8}$ is applied: all  actors have the same opportunity ($o=3$).

\paragraph*{Motivation}
Actors A1, A3 and A4 all benefit from the attack.  Actor A1 conducts these
attacks in order to make money, whereas Actors A3 and A4 are motivated by the
opportunity to cause harm. Thus, Actors A3 and A4's motivation is the same
($m=3$) and higher than that of Actor A1 ($m=1$) since for the latter there are
other ways to make more money faster.

\subsubsection{Attack goal G4}\label{probG4}
\paragraph*{Capacity}
In Scenario $S_{14}$ a replay attack is performed. There is no major challenge
in conducting this attack, which consists in periodically transmitting a Wake-Up
command to the CIED by means of an SDR.  Thus Actors A1, A3 and A4's capacity is
the same ($c=3$). However, in Scenarios $S_{12}$, $S_{11}$ and $S_{4}$, Actor
A1's capacity is higher ($c=3$) than that of A3 and A4 ($c=2$), since these
scenarios consist in implanting a ransomware, and A1 are experts on malicious
code development. For Scenario $S_{15}$ advanced knowledge in computer
programming and architecture is needed. Thus, Actor A3's capacity ($c=3$) will be higher
because they have more human resources and specialized personnel, followed by,
Actors A4 ($c=2$) and A1 ($c=1$).

\paragraph*{Opportunity}
In Scenarios $S_{14}$ and $S_{12}$ there are still constraints in terms of time
and space. The actor must be close to the patient in order to send radio
commands with its antenna to the CIED . Moreover, the attack must take place
while the wireless communication is established in the CIED. As in the other
scenarios where these constraints are presents, the opportunity of Actor A3
($o=2$) is always higher than that of Actors A1 and A4 ($o=1$). In Scenarios
$S_{4}$ and $S_{11}$, the situation is the same, the actor being limited by
space (home or hospital) and time (update sessions).  Normally, when there is
only a space constraint all actors have the same opportunity at the hospital
($S_{11}$) because it is a public place and, while Actor A3 has more opportunity
at home ($S_{4}$). However, since in Scenario $S_{11}$ there is the additional
the time constraint that it happens during an update session, the actors
opportunity will be the same in both scenarios.  Thus, in Scenarios $S_{11}$ and
$S_{4}$ the opportunity of A3 ($o=2$) is higher than that of A1 and A4
($o=1$). In Scenario $S_{15}$, physical access to the targeted system, i.e.~the
monitor, is required. Thus, the opportunity for Actor A3 ($o=2$) is higher than that of
A1 and A4 ($o=1$).

\paragraph*{Motivation}
Actors A1, A3 and A4 all benefit from the attack. By performing these attack
scenarios, A3 ($m=2$) and A4 ($m=3$) would succeed in endangering patients'
lives and consequently harming their quality of life, while A1 ($m=1$) would make
money through ransom.

\subsubsection{Attack goal G5}
\paragraph*{Capacity}
Since there is extensive information about the external programmer behaviour, the
capacity of Actors A3 and A4 ($c=3$) is the same on Scenarios $S_{11(a)}$ and
$S_{11(c)}$ . For Scenario $S_{11(b)}$ knowledge of Cardiology is required, and
Actor A3 is more likely to have access to personnel with such knowlegde or
hiring it.  Thus, A3's capacity ($c=2$) is higher than that of A4 ($c=1$). On
Scenario $S_{12}$, the capacity of the Actors A3 and A4 will be the same
($c=2$). The reasoning is the same as that for Scenario $S_{12}$ (Section~\ref{probG3}),
namely the lack of information concerning the CIED's behaviour and
implementation.

\paragraph*{Opportunity}
The analyis of the opportunity factor for Scenario $S_{14}$
(Section~\ref{probG4}) apply equally to Scenario $S_{12}$. Thus, the opportunity
of A3 ($o=2$) is higher than that of A4 ($o=1$). The same is true for the
analysis of opportunity for Scenario $S_{11}$ on Attack Goal G3
(Section~\ref{probG3}), which applies to Scenarios $S_{11(a)}$, $S_{11(b)}$ and
$S_{11(c)}$. That is to say that the opportunity of A3 ($o=2$) is higher than
that of A4 ($o=1$).

\paragraph*{Motivation}
This attack goal clearly aims at harming the health of an individual. Thus it is Actor A3  ($m=2$) and Actor  A4 ($m=3$) that benefit most from this attack. We do not give them maximum motivation because there are many faster and equally subtle ways to achieve this goal.

\subsubsection{Attack goal G6}
\noindent For Scenario $S_{10}$ the analysis made in Section~\ref{probG2} in terms of
capacity and opportunity equally applies. For Scenarios $S_{11}$ and $S_{14}$,
the capacity and opportunity socres are the same as those of
Section~\ref{probG3} and \ref{probG4},n respectively. In terms of motivation the
same reasoning than for Attack Goal G3 (\ref{probG3}) is applied.\\

\subsection{Combined risk assessment}
\noindent Risk assessment values range between 3 and 48.  They are calculated as the
probability (ranging from 3 to 12) multiplied by the impact (from 1 to 4). We
calculate the risk separately for each impact category. This way of doing things
gives insight of the risk that each threat (scenario, actor) represents
separately for the health, economy, quality of life and privacy impact
categories. Consequently, this analysis responds to the needs of several
different groups such as medical practitioners, regulators, manufacturers and
even patients. Each will know what the riskiest threat is for him and therefore
the one to treat with priority. We ranked the risks in
Table~\ref{tab:RiskCharacterization}.  Depending on the risk value, different
risk management strategies can be chosen and applied. There are four strategies
for managing risk, namely \emph{refuse}, \emph{accept}, \emph{transfer} or
\emph{manage} the risk. The most drastic is of course to refuse the risk, which
is when the risk is considered unacceptable because of the catastrophic
consequences it may have on the victims. In those cases, it is recommended to
prohibit, stop using or remove the system posing the threat. The strategy of accepting the risk is applied when the risk is either
negligible or acceptable. That is to say when the benefits that the system bring
outweigh its potential risks. Transfering the risk relies on giving the risk
management responsibility to a third party such as an insurance company.  This
is a strategy that is not really applicable in those threats where the impact is
on patient health or quality of life. Finally, the risk mitigation or risk
management strategy consists in reducing the risk as much as possible with
available means. This can be done through the updates of the systems, stricter
regulations or even awareness campaigns.

 \begin{table}[]
     \centering
     \begin{tabular}{lll}
         Risk level & Values & Management strategy\\ \hline
          \textbf{\textcolor{red}{--}}  Unacceptable & R=[36,48] & Refuse  \\
           \textbf{\textcolor{orange}{--}} Undesirable & R=[24,35] & Manage  \\
           \textbf{\textcolor{yellow}{--}} Acceptable & R=[12,23] & Accept  \\
           \textbf{\textcolor{green}{--}} Negligeable & R=[3,11] & Accept \\ \hline
     \end{tabular}
     \caption{Risk characterization}
     \label{tab:RiskCharacterization}
 \end{table}

\section{Results and Discussion}
\noindent The attacks goals of inducing medical staff to make errors (G3) and alter device
behavior to endanger patient (G5) represent a risk for patient health. Those to
gain knowledge of device operation and software (G2), induce medical staff to
make errors (G3) and disrupt or lower quality of patient follow-up (G4)
represent an economic risk to manufacturers and health organizations. We can
then note that G3 represents a risk for all groups. In terms of privacy or
degradation of life quality, none of the attack goals represent a potential risk
that needs to be managed. In this section, we focus on those threats
representing either an unacceptable or an undesirable risk for the victims'
health and economy. The risk results of all the threats herein considered can be
found in Table~\ref{tab:RiskCharacterization} of \ref{AppendixA}.

\subsection{Monetary risk assessment}
\subsubsection{Monetary risk assessment by attack goals}

\paragraph*{\textbf{Attack goal G2}}
This attack goal represents a major risk in terms of economic losses. The victim
can be either the manufacturer or the hospital. As hospitals are public
organization, it can be considered that it is the whole society that is the
victim. G2 contains five unacceptable threats (Scenarios $S_4$, $S_5$, $S_6$,
$S_7$ and $S_8$ with all actors). These threats should be managed with high
priority. By analyzing these threats, we can see that the actor's attack
method is always the same, namely exploiting the authentication mechanisms of
the target systems, i.e.~the external devices and cloud-based systems with
which they interact. This fact in itself is good news. On the one hand, external
devices are not constrained by the resource limitations
as the CIED are, so robust authentication solutions can be implemented without significant problems.  There is a plethora of standard robust and proven solutions to secure system authentication, and there is no need to resort to proprietary, unproven solutions.  We, therefore, propose the following solutions.\\

\noindent The threats related to Scenario $S_4$ are solved by securing domestic
networks. To do this, patients must take the habit of securing their network
with a robust password, e.g.~a password containing upper and lower case
characters, numbers and special characters. This password should be periodically
changed. Also, the patient should pay attention to the other Internet of Things
(IOT) devices that are connected to his network, as they can be the entry door
to their network. Accordingly, they should ensure that all devices in their
networks are secured with a password.\\

\noindent To solve the threats associated with Scenario $S_5$, it is essential to insist
that web developers use good code practices and that the source code of web
pages be periodically reviewed.\\

\noindent To mitigate the threats associated with Scenario $S_6$, hospitals and manufacturers should adopt more reliable VPN solutions even if they require more investment. Besides, hospitals and manufacturers should consider recruiting cybersecurity professionals and technical services whose responsibility will be to ensure that there are no cybersecurity threats in their systems and/or networks, including those used for CIED programming and management.\\

\noindent For the threats associated with Scenarios $S_7$ and $S_8$, the solution involves
securing USB ports of monitors and programmers with robust passwords, which
should be continuously modified.\\

\noindent The threat posed by Scenario $S_9$ is not as significant. This means that it
must be managed. The solution is simple: physical security of the targets
devices, in this case, programmers. In addition, it would be necessary to carry
out awareness campaigns among the staff who use those devices, so that they become 
aware of the scope of the problem and therefore more attentive to the physical security of these devices.\\

\paragraph*{\textbf{Attack goal G3}}
\noindent The threats associated with the scenarios $S_{11}$ and $S_{13}$ represent an undesirable risk. In order to mitigate the first threat, hospitals and manufacturers should adopt more reliable VPN solutions. The mitigation of the second threat involves securing the USB ports of the programmers with robust passwords.\\
 
\paragraph*{\textbf{Attack goal G4}}
\noindent The threat related to Scenario $S_{10}$ represents an undesirable risk whose
mitigation is to protect the debugging interface ports with a password.\\

\begin{table*}
\begin{tabular}{ll}
         Risk level &  Management strategy\\ \hline
          \textbf{\textcolor{red}{--}}  Unacceptable& Refuse  \\
           \textbf{\textcolor{orange}{--}} Undesirable & Manage  \\
           \textbf{\textcolor{yellow}{--}} Acceptable & Accept  \\
           \textbf{\textcolor{green}{--}} Negligeable & Accept \\
     \end{tabular}
    
     \centering
     \resizebox{\linewidth}{!}{
       \rowcolors{2}{cyan!10}{cyan!10}
\begin{tabular}{|l l c c|c|c|}
        \hline
Attack goal &Scenario & Attack vector &$P_{rMax}$ &I&R \\\hline
$G_1$ Access patients sensitive data &$S_1$&3&7&1& \cellcolor{green}7\\
 &$S_2$&3&7&1&\cellcolor{green}7\\
 &$S_3$&10,13,14&6&1&\cellcolor{green}6\\
 &$S_4$&15&7&1&\cellcolor{green}7\\ \hline
 $G_{2}$ Gain Knowledge of device operation and software &$S_{4}$&15&9&4&\cellcolor{red}36\\
 &$S_5$&6&11&4&\cellcolor{red}44\\
 &$S_6$&7&9&4&\cellcolor{red}36\\
 &$S_7$&9&10&4&\cellcolor{red}40\\
&$S_{8}$&9&10&4&\cellcolor{red}40\\
 &$S_{9}$&8&7&4&\cellcolor{orange}28\\
 &$S_{10}$&10,11,12&5&4&\cellcolor{yellow}20\\\hline
 $G_{3}$ Induce medical staff to make errors&$S_{11}$&7&8&3&\cellcolor{orange}24\\
&$S_{12}$&1,4,5&7&3&\cellcolor{yellow}21\\
 &$S_{13}$&9&9&3&\cellcolor{orange}27\\\hline
 $G_{4}$ Disrupt or lower quality of patient follow-up&$S_{4}$&15&7&3&\cellcolor{yellow}21\\
 &$S_{11}$&7&7&3&\cellcolor{yellow}21\\
 &$S_{12}$&1,4,5&7&3&\cellcolor{yellow}21\\
 &$S_{14}$&2&7&3&\cellcolor{yellow}21\\
 &$S_{15}$&10&8&3&\cellcolor{orange}24\\\hline
 $G_{5}$ Alter device behavior to endanger patient &$S_{11(a)}$&7&7&3&\cellcolor{yellow}21\\
 &$S_{11(b)}$&7&6&3&\cellcolor{yellow}18\\
 &$S_{11(c)}$&7&7&3&\cellcolor{yellow}21\\
 &$S_{12}$&1,4,5&6&3&\cellcolor{yellow}18\\ \hline
$G_{6}$ Alter device behavior to decrease quality of life &$S_{10}$&10,11,12&7&2&\cellcolor{yellow}14 \\
 &$S_{11}$&7&8&2&\cellcolor{yellow}16\\
 &$S_{14}$&2&8&2&\cellcolor{yellow}16\\ \hline
 \end{tabular} }
        \caption{Results of the monetary risk assessment}
        \label{tabeauDeRisqueMonetaire}
    \end{table*}

\subsubsection{Monetary risk assessment by attack vectors}
\noindent From an economic point of view, the vulnerabilities $V_6$, $V_7$, $V_9$, and $V_{15}$ must be eliminated, because their exploitation constitutes an unacceptable risk for the hospitals and the manufacturers. $V_6$ is eliminated by using good programming practices and revising the source code of the programmers' software. $V_7$ by securing hospital networks, and adopting more reliable VPN solutions.  The security of hospital networks can also be improved by implementing efficient identity and access management (IAM) rules.  For $V_9$, it is necessary to secure the USB ports of the external devices with strong passwords. Finally, securing home networks with strong passwords would eliminate the vulnerability $V_{15}$. Once the vulnerabilities mentioned above have been addressed, vulnerability $V_{8}$ must be managed as a priority because its exploitation constitutes an undesirable risk for hospitals. To do that, they must ensure the physical security of the programmer devices.\\

\subsection{Health risk assessment}
\subsubsection{Health risk assessment by attack goals }
\paragraph*{\textbf{Attack goal G3}}
\noindent The results of Table~\ref{tabeauDeRisqueSante} reveal that G3 is the riskiest attack goal in terms of health. This is because of the unacceptable risk that Scenario $S_{13}$ represents, i.e.~the insertion of malware on the programmer through a USB port connection aimed to generate reading errors.  Among the riskiest threats of this attack goal, this one must be managed with priority.  However, the solution is simple: protect USB port connection with a robust password and frequently change this password.   During our observation of operations in a pacemaker clinic, we observed that it is common practice for staff to record the readings of the programmer (during follow-up sessions) in a USB key and then insert the key into a medical report formatting software in a separate computer system. We recommend that staff pay attention because this USB key could be the target of the actors. They could install the malware on it, and it would infect the programmer. Secondly, the computer where the software is located could also be the target of the actor. This means that the actor could infect the computer, subsequently the computer would infect the USB key, and then the programmer. Thus, it is necessary to pay attention to who is using the USB key, and then to ensure that the computer containing the report formatting software is itself secure (e.g. not connected to the network, unless strictly necessary).\\

\noindent The threats related to Attack Scenarios $S_{11}$ and $S_{12}$ constitute an
undesirable risk that need to be mitigated. For Scenario $S_{11}$, the threat
consists in the insertion of malware into the programmer. $S_{11}$ is achievable
by accessing the device network during the programmer update session. The
threat, as mentioned above, is avoidable by securing the health center network.
Accordingly, it is necessary to implement an efficient method of identity and
access management (IAM) of the computer systems of those entities. On the other
hand, $S_{12}$ threat takes advantage of the improper restriction of
communication channels during the programmer updates. As mentioned in
Section~\ref{Vulnerabilities}, those updates are achieved through a VPN between
the device and the entity in charge of the updates. Thus, the health centers and
manufacturers must invest in reliable solutions of VPN.  For Scenario $S_{12}$,
the threat is the insertion of malware on the CIED. This threat is due to the
lack of robustness of the CIED authentication mechanisms. One potentital
solution consists in implementing more robust authentication mechanisms by using
well-known techniques (e.g.~asymmetric cryptography). However, CIED are limited
in terms of computing resources and such solutions are not the most
appropriate. There are, however, other more adequate solutions, which could be
applied during the CIED manufacturing process. In particular, we propose that
manufacturers use whitelisting techniques in the CIED software, which would
prevent devices other than the programmer from sending commands to the CIED.\\

\paragraph*{\textbf{Attack goal G5}}
As in Attack Goal G3, the achievement of Scenarios $S_{11}$ and $S_{12}$ constitutes undesirable risk that must be managed. The same recommendations made for G3 therefore also apply here. 

\begin{table*}
\begin{tabular}{ll}
         Risk level &  Management strategy\\ \hline
          \textbf{\textcolor{red}{--}}  Unacceptable& Refuse  \\
           \textbf{\textcolor{orange}{--}} Undesirable & Manage  \\
           \textbf{\textcolor{yellow}{--}} Acceptable & Accept  \\
           \textbf{\textcolor{green}{--}} Negligeable & Accept \\
     \end{tabular}
    
     \centering
     \resizebox{\linewidth}{!}{
        \rowcolors{2}{cyan!10}{cyan!10}
\begin{tabular}{|l l c c |c|c|}% 6 columnas
     \hline  
Attack goal &Scenario & Attack vector &$P_{rMax}$ &I&R \\\hline

 $G_{3}$ Induce medical staff to make errors&$S_{11}$&7&8& 4&\cellcolor{orange}32 \\
 &$S_{12}$&1,4,5&7& 4&\cellcolor{orange}28 \\
 &$S_{13}$&9&9& 4&\cellcolor{red}36 \\\hline
 $G_{4}$ Disrupt or lower quality of patient follow-up &$S_{4}$&15&7& 2&\cellcolor{yellow}14\\
 &$S_{11}$&7&7& 2&\cellcolor{yellow}14\\
 &$S_{12}$&1,4,5&7& 2&\cellcolor{yellow}14\\
 &$S_{14}$&2&7& 2&\cellcolor{yellow}14\\
 &$S_{15}$&10&8& 2&\cellcolor{yellow}16\\\hline
 $G_{5}$ Alter device behavior to endanger patient &$S_{11(a)}$&7&7& 4&\cellcolor{orange}28\\
 &$S_{11(b)}$&7&6& 4&\cellcolor{orange}24\\
 &$S_{11(c)}$&7&7& 4&\cellcolor{orange}28\\
 &$S_{12}$&1,4,5&6& 4&\cellcolor{orange}24\\\hline

        \end{tabular}}
        \caption{Results of the health risk assessment}
        \label{tabeauDeRisqueSante}
    \end{table*}
    
\subsubsection{Health risk assessment by attack vectors }
\noindent From a health point of view, vulnerability $V_9$ must be eliminated because its exploitation represents an unacceptable risk to the health of individuals. This is feasible by securing the USB ports of the external devices with strong
passwords. Once $V_9$ is adequately managed,  Vulnerabilities $V_6$, $V_7$ and $V_5$ must be managed as a priority because their exploitation constitutes an undesirable risk. To mitigate the risk that $V_6$ represents, good programming practices and code source revision must take place on the programmer software. To reduce the risk associated with $V_7$, the hospital networks must be secured, and reliable VPN solutions must be applied. Finally, to mitigate $V_5$ it is necessary to apply whitelisting techniques on the CIED.\\

\section{Conclusion}
\noindent As evidenced by previous work, CIED are vulnerable to cyber attacks that use their RF interfaces to communicate with external devices (programmer and home monitor). This fact has been proven by the realization of radio attacks against the CIED RF communication interface in research laboratories \cite{halperin2008pacemakers,marin2016security}. Additionally, the telemetry functionality of the externals devices introduces vectors of cyber attacks \cite{EcosystemER}. Those can include manipulation of the home monitor, interception of transmissions from the home monitor to the cloud and the physician's station, and manipulation of the cloud-based database itself. Although the vulnerabilities mentioned above exist, no attacks have been reported until now in real life, i.e.~in an environment other than the controlled environment of research laboratories.\\

\noindent Thus, it remained to be determined how viable such an attack would be on an actual target (person or device) in the real world. This led us to the following research question: What are the real risks of cyber attacks onto  CIED and the systems they depend on (programmer, monitor, cloud-based systems)? To answer this question,  we carried out a realistic risk analysis of such attacks, with regards to their impact at four scales: health, economy, quality of life and privacy. We proceeded in this way because the problem under study affects many different groups namely: patients,  practitioners, manufacturers, and more broadly states.  Accordingly, separating the scales aims to individually support those  groups objectives in terms of risk management.\\

\noindent We did three kinds of analysis.  First, an actor-based risk analysis to determine who the actors are and what their attack goals are. This analysis allowed us to determine the level of impact of the attacks. We then made a scenario-based risk analysis to determine the probability of occurrence of the attacks. Finally, we performed a combined risk analysis by considering the impact and probability results. We determined the most dangerous attack goals on the one hand, and the most dangerous vulnerabilities on the other.\\
 
\noindent Our work reveals that the vulnerabilities associated with the RF communication interface of CIED represents an acceptable risk. This is due to the fact that these vulnerabilities have a low probability of being successfully exploited in real conditions (environment other than a research laboratory). However, the network and Internet connectivity of external devices represents a risk that in some cases is unacceptable, i.e.~a risk that must be absolutely refused. The answer to our research question is therefore that the real risk is in the external devices and not in the CIED and that this risk is due to the increasing connectivity of said devices.  We can therefore see that the problem under study is the medical variant of the trendy cyber-security problem: the lack of security of connected objects (Internet of Thing or IOT).\\

\noindent Indeed, among the 15 vulnerabilities identified, four constitute an unacceptable
risk. They are $V_6$, $V_7$, $V_9$, and $V_{15}$ and are all to external
devices. Five other vulnerabilities ($V_1$, $V_4$, $V_5$, $V_8$, and$V_{10}$)
represent an undesirable risk, i.e.~a risk that must be addressed. Among the
latter two are vulnerabilities specific to CIED ($V_1$,$V_4$). However, their
exploitation will have an impact if and only if another specific programmer
vulnerability is successfully exploited ($V_5$). There are already existing
solutions to avoid all those vulnerabilities. The parties involved have to put
them into practice. In order to achieve this, stronger regulation and
legislation is needed. These should not be limited to good practice guidelines
without any force of law as is the case today. In the same manner, the FDA or
Health Canada (or other similar national organizations) should impose that
hardware components of medical equipment pass a set of certification tests
including cyber security assessment in order to be accepted in the market; the
same should be the case for their software components. Finally, the various
involved parties (practionners, patients, etc.)  should be duly informed of the
origin, nature and scope of the threats and how to protect themselves at their
level. This information must be disclosed in language that is understandable to
them so that they can take part in the solution.\\

\noindent Moreover, our analysis revealed that the attack goals (G2) \emph{Gain knowledge of device operation and software} and (G3) \emph{Induce medical staff to make errors} are the main attack goals of the actors.  This result shows that while attacks on these devices affect patients, the patients are not always the target as we may have thought so far. The targets in many cases are manufacturers (intellectual property theft) and practitioners (threat of civil liability) for purely economic reasons. Manufacturers should, therefore, be aware of the problem and focus on the computer security of their equipment.  The first step to this is avoiding secrecy regarding the software and architecture of their equipment. As has been often posited, code is more secure when it is open source since several people can test it and report errors so that they can be patched. This secrecy about code instead of protecting manufacturers, exposes them more to cyber security risk. Health centers have to become more selective and demanding with the equipment they buy and implant on patients, as this would allow them to put more pressure on manufacturers to make the right cyber security choices.\\

\newpage
\appendix
\section{Risk assessment by attack goals and impact type}
\label{AppendixA}

\begin{table*}[h!] 
    \begin{tabular}{ll}
        \hspace{0.6\linewidth}  \textbf{\textcolor{red}{---}} & Unacceptable risk  \\
        \hspace{0.6\linewidth}   \textbf{\textcolor{orange}{---}}& Undesirable risk  \\
        \hspace{0.6\linewidth}   \textbf{\textcolor{yellow}{---}}& Acceptable risk   \\
         \hspace{0.6\linewidth}  \textbf{\textcolor{green}{---}}& Negligeable risk  \\
     \end{tabular}
     \bigskip
        \rowcolors{3}{cyan!10}{cyan!10}
\begin{tabular}{| c c c c |c|c|c|c|c|c|c|c|}% 9 columnas
        \hline
          & & &  & \multicolumn{2}{c|}{H}&\multicolumn{2}{c|}{M}&\multicolumn{2}{c|}{LQ}&\multicolumn{2}{c|}{P}\\
Attack goal &Scenario & Attack vector &$P_{rMax}$ &I&R &I&R &I&R  &I&R\\\hline
$G_1$ &$S_1$&3&7& -&- &1& \cellcolor{green}7 &-&-  &2&\cellcolor{yellow}14\\
 &$S_2$&3&7& -&- &1&\cellcolor{green}7 &-&-  &2&\cellcolor{yellow}14\\
 &$S_3$&10,13,14&6& -&- &1&\cellcolor{green}6 &-&-  &2&\cellcolor{yellow}12\\
 &$S_4$&15&7& -&- &1&\cellcolor{green}7 &-&-  &2&\cellcolor{yellow}14\\ \hline
 $G_2$ &$S_{4}$&15&9& -&- &4&\cellcolor{red}36 &-&-  &-&-\\
 &$S_5$&6&11& -&- &4&\cellcolor{red}44 &-&-  &-&-\\
 &$S_6$&7&9& -&- &4&\cellcolor{red}36 &-&-  &-&-\\
 &$S_7$&9&10& -&- &4&\cellcolor{red}40 &-&-  &-&-\\
 &$S_{8}$&9&10& -&- &4&\cellcolor{red}40 &-&-  &-&-\\
 &$S_{9}$&8&7& -&- &4&\cellcolor{orange}28 &-&-  &-&-\\
 &$S_{10}$&10,11,12&5& -&- &4&\cellcolor{yellow}20 &-&-  &-&-\\\hline
 $G_3$ &$S_{11}$&7&8& 4&\cellcolor{orange}32 &3&\cellcolor{orange}24 &1&\cellcolor{green}8  &-&-\\
 &$S_{12}$&1,4,5&7& 4&\cellcolor{orange}28 &3&\cellcolor{yellow}21 &1&\cellcolor{green}7  &-&-\\
 &$S_{13}$&9&9& 4&\cellcolor{red}36 &3&\cellcolor{orange}27 &1&\cellcolor{green}9  &-&-\\\hline
 $G_4$ &$S_{4}$&15&7& 2&\cellcolor{yellow}14 &3&\cellcolor{yellow}21 &1&\cellcolor{green}7  &-&-\\
  &$S_{11}$&7&7& 2&\cellcolor{yellow}14 &3&\cellcolor{yellow}21 &1&\cellcolor{green}7  &-&-\\
 &$S_{12}$&1,4,5&7& 2&\cellcolor{yellow}14 &3&\cellcolor{yellow}21 &1&\cellcolor{green}7  &-&-\\
 
 &$S_{14}$&2&7& 2&\cellcolor{yellow}14 &3&\cellcolor{yellow}21 &1&\cellcolor{green}7  &-&-\\

 &$S_{15}$&10&8& 2&\cellcolor{yellow}16 &3&\cellcolor{orange}24 &1&\cellcolor{green}8  &-&-\\
  \hline
 $G_{5}$ &$S_{11(a)}$&7&7& 4&\cellcolor{orange}28 &3&\cellcolor{yellow}21 &-&-  &-&-\\
 &$S_{11(b)}$&7&6& 4&\cellcolor{orange}24 &3&\cellcolor{yellow}18 &-&-  &-&-\\
 &$S_{11(c)}$&7&7& 4&\cellcolor{orange}28 &3&\cellcolor{yellow}21 &-&-  &-&-\\
 &$S_{12}$&1,4,5&6& 4&\cellcolor{orange}24 &3&\cellcolor{yellow}18 &-&-  &-&-\\ \hline

$G_{6}$ &$S_{10}$&10,11,12&7& -&- &2&\cellcolor{yellow}14 &2&\cellcolor{yellow}14  &-&-\\
% &$S_{26}$&7&7& -&- &2&14 &2&14  &-&-\\
 &$S_{11}$&7&8& -&- &2&\cellcolor{yellow}16 &2&\cellcolor{yellow}16  &-&-\\
  &$S_{14}$&2&8& -&- &2&\cellcolor{yellow}16 &2&\cellcolor{yellow}16  &-&-\\
 \hline
        \end{tabular}
        \caption{Risk assessment results}
        \label{tabeauDeRisque}
   \end{table*}

\newpage
\section{Sequence of events of the attack scenarios}\label{AppendixB}
\noindent$S_{1}$ : Radio attack on the CIED-Programmer wireless communications.\\
($e_{1}$)Acquire the hardware (SDR, antenna, signal processing software)\\
($e_{2}$)Go to the hospital\\
($e_{3}$)Be located at a distance relatively close to the CIED\\
($e_{4}$)Configure the SDR in reception mode\\
($e_{5}$) Perform a frequency scan of the MICS band to determine the transmission frequency of the CIED\\
($e_{6}$) Intercept and record the signal transmitted by the CIED\\
($e_{7}$)Read the patient’s health data ($V_{3}$)\\

\noindent$S_{2}$ : Radio attack on the CIED-Monitor wireless communications.\\
($e_{1}$)Acquire the hardware (SDR, antenna, signal processing software).\\
($e_{2}$)Go to the patient's home\\
($e_{3}$)Be located at a distance relatively close to the CIED\\
($e_{4}$)Configure the SDR in reception mode\\
($e_{5}$) Perform a frequency scan of the MICS band to determine the transmission frequency of the CIED\\
($e_{6}$)Intercept and record the signal transmitted by the CIED\\
($e_{7}$) Read the patient’s health data ($V_{3}$)\\

\noindent$S_{3}$: Unauthorized physical access to the monitor content\\
--------------------Using the JTAG interface---------------------------\\
($e_{1}$)Acquire the hardware (F to F jumper wire, in-debugger-circuits, PC with IDLE debugger)\\
($e_{1}$)Go to the patient's home\\
($e_{2}$)Take the patient's monitor\\
($e_{3}$)Connect one extremity of the F to F jumper wire to the monitor debug port (exploiting $V_{10}$)\\
($e_{4}$)Connect the other extremity of the F to F jumper wire to the in-debugger-circuits\\
($e_{5}$)Connect the in-debugger-circuit to the PC\\
($e_{6}$)Access the monitor memory by means of the IDLE debugger\\
($e_{7}$)Use $V_{13}$ and $V_{14}$ to adjust the server settings and credentials to authenticate to them\\
($e_{8}$)Access the server by means of the information obtained in ($e_{8}$)\\
($e_{9}$)Read the patient's medical data\\

--------------------Using the UART interface---------------------------\\
($e_{1}$)Acquire the hardware (Pirate bus, PC with IDLE debugger)\\
($e_{2}$)Go to the patient's home\\
($e_{3}$)Take the patient's monitor\\
($e_{4}$)Connect one end of the pirate bus to the monitor debug port (exploiting $V_{10}$)\\
($e_{5}$)Connect the other pirate bus end to the PC containing the IDLE debugger\\
($e_{6}$)Access the monitor memory by means of the IDLE debugger\\
($e_{7}$)Use $V_{13}$ and $V_{14}$ to adjust the server settings and credentials to authenticate to them\\
($e_{8}$)Access the server by means of the information obtained in ($e_{7}$)\\
($e_{9}$)Read the patient's medical data\\

\noindent$S_{4}$ : Network attack on the Monitor\\
($e_{1}$)Gain access to the patient’s router  the day of the monitor’s update\\
($e_{2}$)Intercept the updated firmware ($V_{15}$)\\
($e_{3}$)Replace the firmware with a backdoor\\

\noindent$S_{5}$: Web attack on programmers' SW deployment network server\\
($e_{1}$)Find the URL in which the programmer (app) retrieve files from the server\\
($e_{2}$)Modify URL with commands and web server escape code\\
($e_{3}$)Send the URL to the server(via http request) ($e_{3}$)\\
($e_{4}$)Extract the desired files\\

\noindent$S_{6}$ :Network attack on the programmer’s\\
($e_{1}$)Go to the hospital the day of the update\\
($e_{2}$)Access the programmer’s network\\
($e_{3}$)Leverage $V_{7}$ to gain access to the programmer\\
($e_{3}$)Extract the desired files\\

\noindent$S_{7}$: Network attack on the Monitor\\
($e_{1}$)Go to the patient home\\
($e_{2}$)Acces the patient network\\
($e_{3}$)Acces the monitor’s USB port ($V_{9}$)\\
($e_{4}$)Navigate in the file system and extract the desired files\\

\noindent$S_{8}$: Network attack on the Programmer\\
($e_{1}$)Go to the hospital\\
($e_{2}$)Acces the hospital network\\
($e_{3}$)Acces the monitor’s USB port ($V_{9}$)\\
($e_{4}$)Navigate the file system and extract the desired files\\

\noindent$S_{9}$: Physical attack on the Programmer\\
($e_{1}$)Go to the hospital\\
($e_{2}$)Extract the programmer’s removable hard drive($V_{8}$)\\

\noindent$S_{10}$: Physical attack on the monitor\\
--------------------Using the JTAG interface---------------------------\\
($e_{1}$)Acquire the hardware ( F to F jumper wire, in-debugger-circuits, PC with IDLEs debugger)\\
($e_{2}$)Go to the patient’s home\\
($e_{3}$)Take the patient's monitor\\
($e_{4}$)Connect one end of the F to F jumper wire to the monitor debug port ($V_{10}$)\\
($e_{5}$)Connect the other end of the F to F jumper wire to the in-debugger-circuits\\
($e_{6}$)Connect the in-debugger-circuit to the PC\\
($e_{7}$)Access the monitor memory by means of the IDLE debugger\\
($e_{8}$)Use $V_{11}$ and $V_{12}$ to adquer the credentials of OS\\
($e_{9}$)Access the OS of the monitor by means of the information obtained in $e_{8}$\\
($e_{10}$)Read the OS\\
--------------------Using the UART interface---------------------------\\
($e_{1}$)Acquire the hardware (Pirate bus, PC with IDLE debugger)\\
($e_{2}$)Go to the patient’s home\\
($e_{3}$)Take the patient's monitor\\
($e_{4}$)Connect one end of the pirate bus to the monitor debug port (V10)\\
($e_{5}$)Connect the other pirate bus end to the PC containing the IDLE debugger.\\
($e_{6}$)Access the monitor memory by means of the IDLE debugger\\
($e_{7}$)Use $V_{11}$ and $V_{12}$ to acquire the credentials of the OS\\
($e_{8}$)Access the OS of the monitor by means of the information obtained in $e_{7}$\\
($e_{9}$)Read information about OS\\

\noindent$S_{11}$: Network attack on the programmer\\
($e_{1}$)Gain access to the CIED room consultation the day of the update\\
($e_{2}$)Intercept the updated firmware ($V_{7}$)\\
($e_{3}$)Replacing the firmware with malware\\

\noindent$S_{12}$: Radio attack on the CIED \\
($e_{1}$)Acquire the hardware (SDR, antenna, signal processing software)\\
($e_{2}$)Go to the hospital\\
($e_{3}$)Be located at a distance relatively close to the CIED\\
($e_{4}$)Configure the SDR in Transmission mode\\
($e_{5}$) Perform a frequency scan of the MICS band to determine the Programmer's transmission frequency\\
($e_{6}$) Transmit commands (via RF signals) to the CIED ($V_{1}$,$V_{4}$,$V_{5}$)\\

\noindent$S_{13}$: Network attack on the programmer\\
($e_{1}$)Go to the hospital\\
($e_{2}$)Access the hospital network\\
($e_{3}$)Access the monitor’s USB port($V_{9}$)\\
($e_{4}$)Insert a malware\\

\noindent$S_{14}$: Radio attack on the CIED \\
($e_{1}$)Acquire the hardware (SDR, antenna, signal processing software)\\
($e_{2}$)Go to the hospital\\
    ($e_{3}$)Be located at a distance relatively close to the CIED\\
($e_{4}$)Configure the SDR in Transmission mode\\
($e_{5}$) Perform a frequency scan of the MICS band to determine the Programmer's transmission frequency\\
($e_{6}$) Transmit Wake-up commands (via RF signals) to the CIED periodically ($V_{1}$,$V_{2}$,$V_{4}$)\\



%% References


\bibliographystyle{model1-num-names}
\bibliography{Document}






%\end{thebibliography}
\newpage 
\section*{VITAE}
\noindent \textbf{Mikaela Ngambo\'{e}} holds a degree in Telecommunications Engineering from the Universidad Polit\'ecnica de Madrid. She is currently a Master's student at the \'{E}cole Polytechnique de Montr\'{e}al, within the Computer Systems Security Laboratory.  Her research has focused on determining the risk factors that increase the exposure of cyber physical systems to computer attacks.  She has studied this issue in systems ranging from traditional information and communication technologies to implantable medical devices.  

\medskip
\noindent \textbf{Paul Berthier} holds a degree in Telecommunications Engineering from the \'{E}cole nationale sup\'{e}rieure des T\'{e}l\'{e}communications. He holds a M.A.Sc.\ from the \'{E}cole polytechnique de Montr\'{e}al.  His master's project focused on the cyber security of the  Automatic Dependant Surveillance-Broadcast protocol (ADS-B) that allows aircraft to report position, altitude and speed of in real time, and which is intended to replace radar in civil aviation applications.  Mr. Berthier worked for two years as a Research Associate in the Computer Systems Security Laboratory at the \'Ecole Polytechnique de Montr\'{e}al, managing the aviation cyber security research team in that organization. He works as Senior Cybersecurity Advisor in a private company.  
\medskip

\noindent \textbf{Nader Ammari} holds an engineering degree from the  \'{E}cole Nationale des Sciences de l'Informatique (ENSI). He holds a M.A.Sc.\ from the \'{E}cole Polytechnique de Montr\'{e}al.  His master's project focused on the improvement of defenses mechanism of the Return Oriented Programming (ROP) based attacks. An attack method by which it is possible to modify the behavior of a program without injecting malicious code on it. Mr. Ammari worked for one year as the IT security engineer of a firm. He is a Research Associate in the Computer Systems Security Laboratory at the \'Ecole Polytechnique de Montr\'{e}al.
\medskip

\noindent \textbf{Dr.~Katia Dyrda} holds degrees in Engineering Physics from Queen's University and a degree in Medicine from the University of Ottawa.  She trained in internal medicine at Queen's University and then specialized in adult cardiology at the Université de Montréal. Following this, she trained in electrophysiology at the Montreal Heart Institute (MHI) and at the Leiden University Heart Centre in The Netherlands. She now specializes in complex ablation at the MHI where she has been since 2013. She also serves as Cardiology residency program site director for the MHI and EP training program director for Université de Montreal. Throughout, she has maintained her ties to engineering through multiple research projects but with a special interest in electromagnetic interference with cardiac implantable devices.

\medskip
\noindent \textbf{Dr.~Jos\'{e} M.\ Fernandez, Eng.,} is a Full Professor in the Computer \& Software Engineering Department at the \'{E}cole Polytechnique de Montr\'{e}al, where he leads the Computer Systems Security Laboratory. His research interests include security of cyber physical systems, in particular critical infrastructure, industrial control systems, and transportation systems in aviation and vehicular traffic.
Dr. Fernandez holds a Ph.D.\ from the Universit\'{e} de Montréal, an M.A.Sc.\ from the University of Toronto and two B.Sc.\ degrees from the Massachusetts Institute of Technology (MIT). Apart from its academic activity, he has several years of professional experience in computer security in the private and public sector.

\end{document}